\documentclass{emulateapj}
\usepackage{amssymb,amsmath,amsthm}
\usepackage{natbib}
\usepackage{longtable,url}

\newcommand{\chandra}{\it Chandra\rm}

\newcommand{\clusterB}{Abell~2204}
\newcommand{\clusterA}{Abell~2631}

\newcommand{\ho}{\mbox{$H_{\mbox{\tiny 0}}$}}
\newcommand{\da}{\mbox{$D_{\mbox{\tiny A}}$}}
\newcommand{\Sx}{\mbox{$S_{\mbox{\tiny X}}$}}
\newcommand{\Tsz}{\mbox{$T_{\mbox{\tiny SZ}}$}}

\newcommand{\sigT}{\mbox{$\sigma_{\mbox{\tiny T}}$}}
\newcommand{\Tcmb}{\mbox{$T_{\mbox{\tiny CMB}}$}}

\newcommand{\Lamee}{\mbox{$\Lambda_{ee}$}}

\newcommand{\Mgas}{\mbox{$M_{\mbox{\scriptsize gas}}$}}
\newcommand{\Mtot}{\mbox{$M_{\mbox{\scriptsize tot}}$}}

\newcommand{\fgas}{\mbox{$f_{\mbox{\scriptsize gas}}$}}

\newcommand{\LCDM}{\mbox{$\Lambda$CDM}}
\newcommand{\Omegak}{\mbox{$\Omega_{\mbox{\scriptsize k}}$}}
\newcommand{\OmegaM}{\mbox{$\Omega_{\mbox{\scriptsize M}}$}}
\newcommand{\OmegaL}{\mbox{$\Omega_\Lambda$}}

\begin{document}
\title{Joint analysis of X-ray and Sunyaev Zel'dovich observations of galaxy clusters using an
analytic model of the intra-cluster medium}

\author{Nicole~Hasler,\altaffilmark{1} 
Esra~Bulbul,\altaffilmark{1} 
Massimiliano~Bonamente,\altaffilmark{1,2} 
John~E.~Carlstrom,\altaffilmark{3,4,5,6}
Thomas~L.~Culverhouse,\altaffilmark{3,4}
Megan~Gralla,\altaffilmark{3,4}
Christopher~Greer,\altaffilmark{3,4}
David~Hawkins,\altaffilmark{7}
Ryan~Hennessy,\altaffilmark{3,4}
Marshall~Joy,\altaffilmark{2}
Jeffery~Kolodziejczak,\altaffilmark{2}
James~W.~Lamb,\altaffilmark{7}
David~Landry,\altaffilmark{1}
Erik~M.~Leitch,\altaffilmark{3,4}
Adam~Mantz,\altaffilmark{3,4,8}
Daniel~P.~Marrone,\altaffilmark{3,9}
Amber~Miller,\altaffilmark{10,11}
Tony~Mroczkowski,\altaffilmark{12,13}
Stephen~Muchovej,\altaffilmark{7}
Thomas~Plagge,\altaffilmark{3,4}
Clem~Pryke,\altaffilmark{3,4,5}
and David~Woody\altaffilmark{7}
}

\altaffiltext{1}{Department of Physics, University of Alabama, Huntsville, AL 35899}
\altaffiltext{2}{Space Science-VP62, NASA Marshall Space Flight Center, Huntsville, AL 35812}
\altaffiltext{3}{Kavli Institute for Cosmological Physics, University of Chicago, Chicago, IL 60637}
\altaffiltext{4}{Department of Astronomy and Astrophysics, University of Chicago, Chicago, IL 60637}
\altaffiltext{5}{Enrico Fermi Institute, University of Chicago, Chicago, IL 60637}
\altaffiltext{6}{Department of Physics, University of Chicago, Chicago, IL 60637}
\altaffiltext{7}{Owens Valley Radio Observatory, California Institute of Technology, Big Pine, CA 93513}
\altaffiltext{8}{NASA Goddard Space Flight Center, Greenbelt, MD 20771}
\altaffiltext{9}{Hubble Postdoctoral Fellow}
\altaffiltext{10}{Columbia Astrophysics Laboratory, Columbia University, New York, NY 10027}
\altaffiltext{11}{Department of Physics, Columbia University, New York, NY 10027}
\altaffiltext{12}{Einstein Postdoctoral Fellow}
\altaffiltext{13}{Department of Physics and Astronomy, University of Pennsylvania, Philadelphia, PA 19104}

\begin{abstract}
We perform a joint analysis of X-ray and Sunyaev Zel'dovich (SZ) effect data
using an analytic model that describes the gas properties of galaxy clusters.
The joint analysis allows the measurement of the cluster gas mass fraction profile and Hubble constant
independent of cosmological parameters.  Weak cosmological priors are used to calculate
the overdensity radius within which the gas mass fractions are reported.  Such an analysis
can provide direct constraints on the evolution of the cluster gas mass fraction with redshift.
We validate the model and the joint analysis on high signal-to-noise data from the \chandra\ X-ray
Observatory and the Sunyaev-Zel'dovich Array for two clusters, Abell~2631 and  Abell~2204.
\end{abstract}


\section{Introduction}
Galaxy clusters trace the growth of structure in the universe.  Their abundance
and evolution is critically sensitive to underlying cosmological parameters 
such as $\Omega_{M}$, $\sigma_{8}$,
and the dark energy equation of state parameter $w$.
Recent work has focused on using galaxy clusters to constrain cosmology,
including dark energy constraints from X-ray and joint X-ray/SZ measurements of the 
gas mass fraction \citep{allen2008,ettori2009,laroque2006},
cosmological parameter constraints from the growth of structure via
X-ray \citep{mantz2008,mantz2010,vikhlinin2009} 
and SZ cluster surveys \citep{vanderlinde2010,marriage2010,sehgal2010,muchovej2011,williamson2011,benson2011}.

In this paper, we present a method for the joint analysis of X-ray and SZ cluster observations using a self-consistent 
analytic model for the physical properties of the intra-cluster medium \citep{bulbul2010}. The model
provides analytic expressions for the radial density, temperature, and pressure profiles, and 
is therefore simultaneously applicable to both X-ray and SZ observables.
The joint analysis allows measurement of the cluster
gas mass fraction without the need to impose external priors on cosmological
parameters such as the Hubble expansion rate $H(z)$.
Such an analysis applied to a sample of clusters can directly probe the evolution of cluster gas mass fractions with redshift.
We demonstrate the method using high signal-to-noise data from 
\chandra\ and Sunyaev-Zel'dovich Array (SZA) observations of two clusters, \clusterA\ and \clusterB.

The method developed in this paper combines all available Chandra
X-ray data (both imaging and spectroscopic) with SZ observations,
using these to determine the angular diameter distance and cluster
mass with minimal cosmological assumptions.
We chose the \cite{bulbul2010} model for this analysis since it
describes the three thermodynamical cluster properties -- density,
temperature, and pressure -- with a consistent set of parameters that
are both readily interpreted. 
The temperature profile linking density and
pressure is both easily calculated and observationally motivated.
This work differs from  \cite{mroczkowski2009} who used SZ and X-ray imaging data only 
 with a simplified, core-cut form
of the \cite{vikhlinin2006} model to describe the X-ray density and
the \cite{nagai2007b} parameterization of the 
SZ pressure,
and for which the inferred
temperature profile did not reduce to a compact,
accessible expression.

This paper is structured as follows: 
Section \ref{sec:reduction} describes the data reduction and analysis, 
Section \ref{sec:joint} the modeling of X-ray and SZ data, and
Section \ref{sec:app}  the joint analysis of the X-ray and SZ data. 
We present and discuss our conclusions in Section \ref{sec:disconc}.

\begin{figure*}[t!]
\begin{center}
\includegraphics[angle=-90,width=3in]{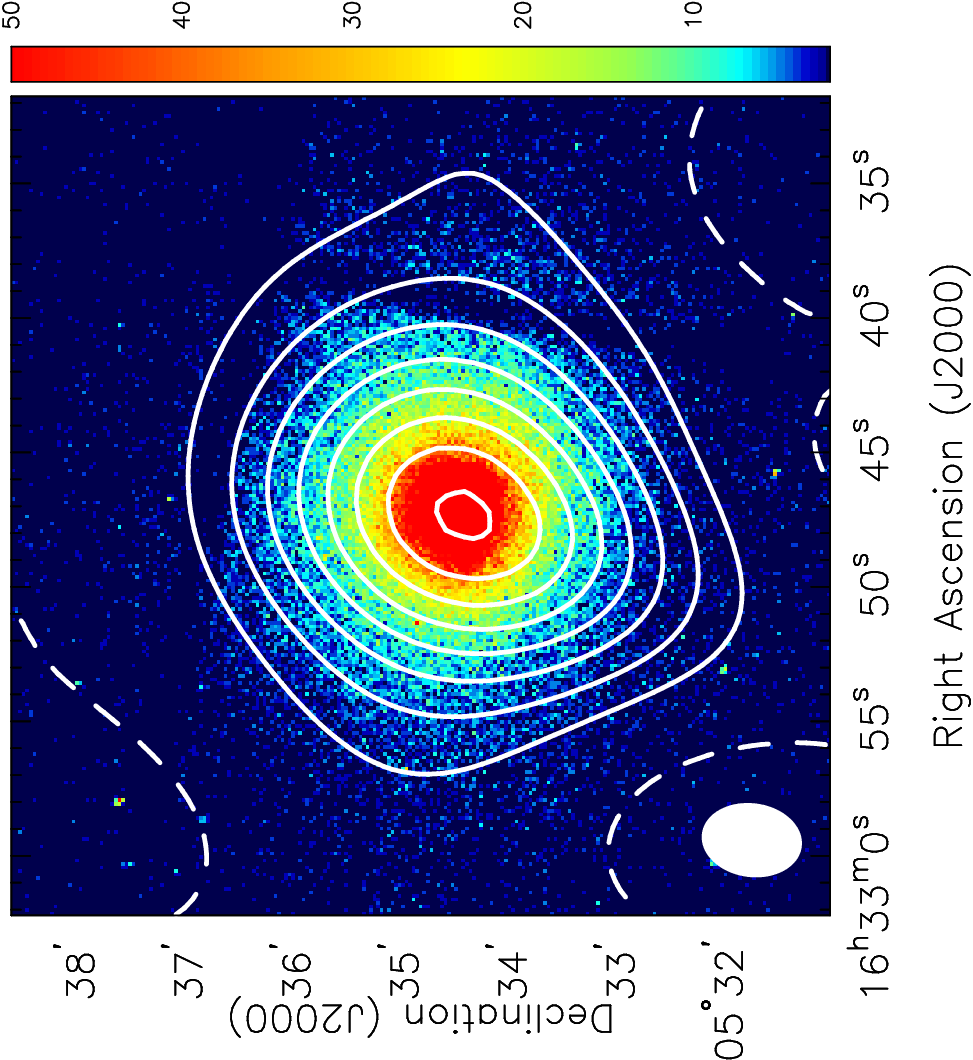}
\includegraphics[angle=-90,width=3in]{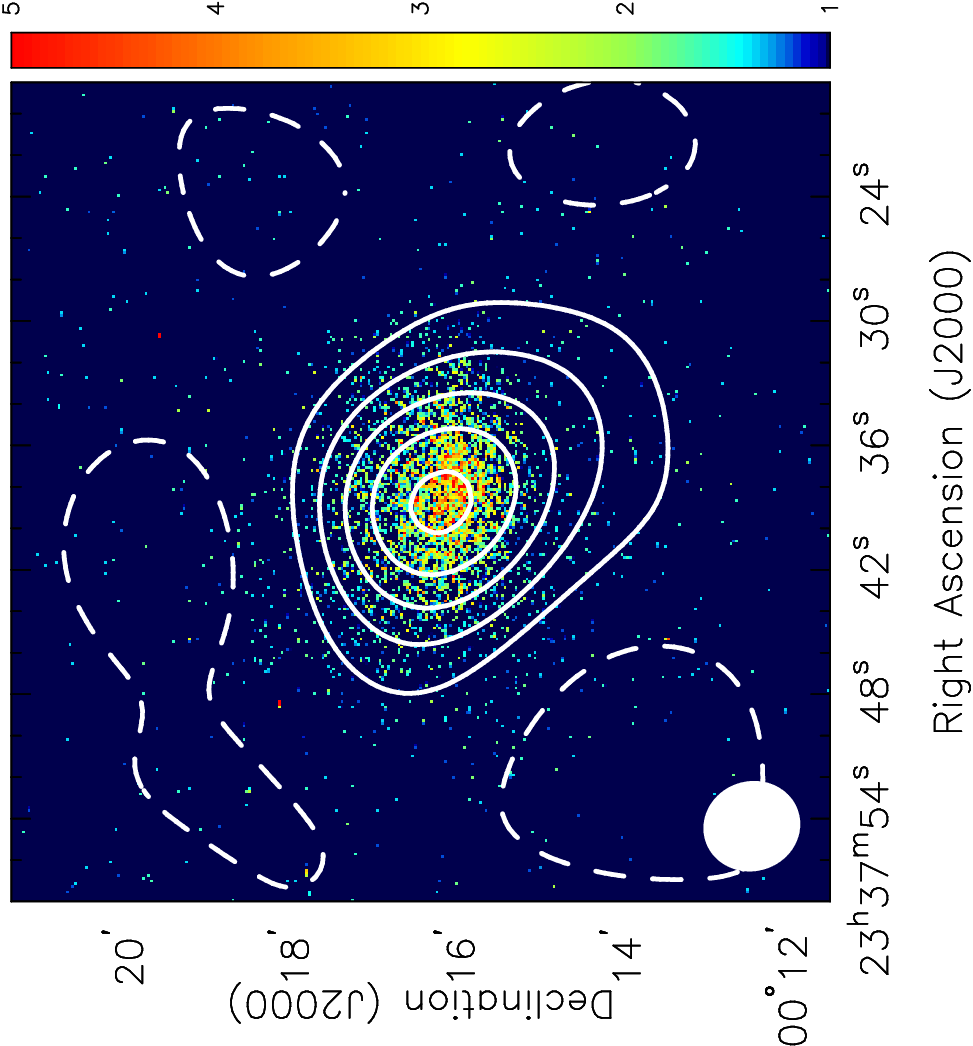}
\end{center}
\caption{SZ contours overlaid on X-ray false color images 
of Abell 2631 ({\it left}) and
Abell 2204 ({\it right }).
The \chandra\ X-ray surface brightness data are from
the energy range 0.7--7.0~keV.  The color bars reflect the number of counts detected by Chandra
in the 0.7-7~keV band, with a pixel size of 1.97 arcsec.
The SZ data are from the SZA, and the contour levels are
(+2,-2,-4,-6,-8,...) times the rms noise (see Table \ref{tab:info}).  
The FWHM of the synthesized beams for these SZ
observations are shown in the lower left corner of each image. 
}
\label{fig:xsz}
\end{figure*}

\section{Data Reduction}
\label{sec:reduction}

We selected  a non-cool core cluster
(Abell~2631) and a cool-core cluster (Abell~2204) with high quality
X-ray and SZ observations to demonstrate the method of analysis.

\subsection{\chandra\ Imaging and Spectroscopy}

The \chandra\ X-ray data are in the form of event files, which we 
use to generate both images and spectra.  
Additional blank-sky composite event files are used for 
background subtraction.  The event files are reduced 
using CIAO 4.3.1 and CALDB 4.3, following the
reduction procedure described in \cite{bulbul2010}.
Details of each cluster observation can be found in Table 1.

The X-ray images in the 0.7-7 keV band are used to measure the X-ray surface brightness profile of the cluster.
To subtract the background from the surface brightness, we rescale the blank-sky image
to match the cluster surface brightness in a peripheral region that is
free of cluster signal. The peripheral regions we chose are at a distance $\geq$ 500 arcsec 
from the cluster center for Abell~2631 (corresponding to 2.7 Mpc in
the standard flat $\Lambda$CDM cosmology) and $\geq$ 550 arcsec (1.7 Mpc) for Abell~2204.

Spectra are extracted in annular regions centered at the peak of the X-ray
emission. These regions cover an area out to the radius
where the surface brightness profile reaches the background, 
which is near $r_{500}$ (the radius within which the average density is
$500$ times the critical density at the cluster redshift) for these 
observations.
From the blank-sky data, background spectra are also extracted and processed.
We then rescale the blank-sky spectra to match the count rate of the cluster 
spectra in the 9.5-12~keV band. In this band, \chandra\ has no effective area for the detection
of photons, and the detected counts originate from a particle background
that is time variable. \cite{hickox2006} showed that while the flux within the
2--7~keV and 9.5--12~keV energy bands can vary with time, the ratio of the
two bands remains constant.  Subtracting the blank-sky data rescaled by
the higher-energy band therefore accurately removes the background 
from the lower-energy band.

After rescaling the blank-sky spectra and removing the background from 
our cluster data, residuals may still be present in the soft 0.7--2~keV
energy band. These soft X-ray residuals may be due to Galactic and extragalactic emission,
and may vary as function of position \citep[e.g.,][]{snowden1997} and 
time \citep[e.g.,][]{takei2008}.
For each cluster observation, we use a peripheral region that is free of
cluster emission---the same region used to rescale the background 
images---to determine whether soft residuals are present after the 
blank-sky background has been subtracted.  We detect the presence of soft 
residuals in both clusters. 
The residual spectra are
fit using a phenomenological model that includes a power law and 
a plasma emission model, 
and this model is rescaled by area and included
in the spectral fit for each annulus \citep[e.g.,][]{snowden1998,nevalainen2005,maughan2008}.

\subsection{Interferometric Observations with the SZA}
The two clusters were observed with the Sunyaev-Zel'dovich Array (SZA), an eight-element 
interferometer designed for the detection and imaging of the SZ effect.
Each antenna in the array is 3.5~m in diameter and has a primary beam 
FWHM of 10.7$^\prime$ at the center frequency of the observing band (31~GHz).
For these observations, six antennas were closely
packed together to provide sensitivity to arcminute-scale SZ
signals, and the remaining two antennas were placed further out to
constrain the flux contributions from unresolved radio sources, as described in \citet{muchovej2007}.
The unflagged on-source time for Abell~2631 was 16.1~hours and 19.6~hours for Abell~2204.
The details of the observations are given in Table~1,
and the radio sources detected in each field are listed in Table 2.
In the analysis of the cluster SZ effect described below, the parameters of the 
SZ decrement and the radio sources are fit simultaneously.

The SZA data are reduced using a set of routines written in MATLAB\footnote{http://www.mathworks.com/products/matlab}
that constitute a complete pipeline for flagging, calibrating, and reducing
visibility data.  The reduction pipeline, described 
in \citet{muchovej2007}, converts the data to physical units and 
corrects for instrumental phase and amplitude variations.  Data 
are flagged for corruption due to bad weather, sources of radio 
interference, and other instrumental effects that 
could impact their quality.  The pipeline outputs calibrated unflagged visibilties, i.e., components of the 
Fourier transform of the sky brightness multiplied by the primary beam response, along with their corresponding
statistical weights and positions in the Fourier ($u-v$) plane.

\begin{figure*}[ht!]
\centering
\includegraphics[angle=-90,width=3in]{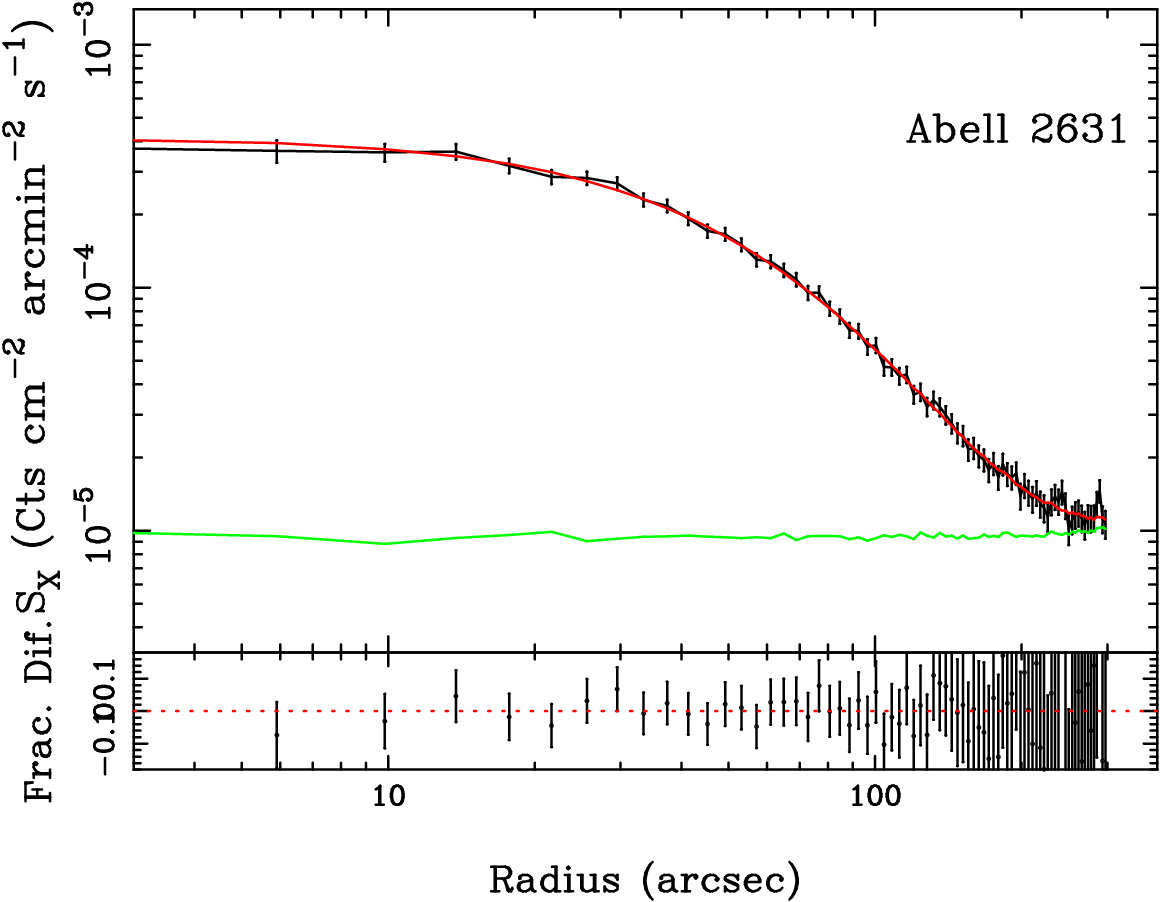}
\includegraphics[angle=-90,width=3in]{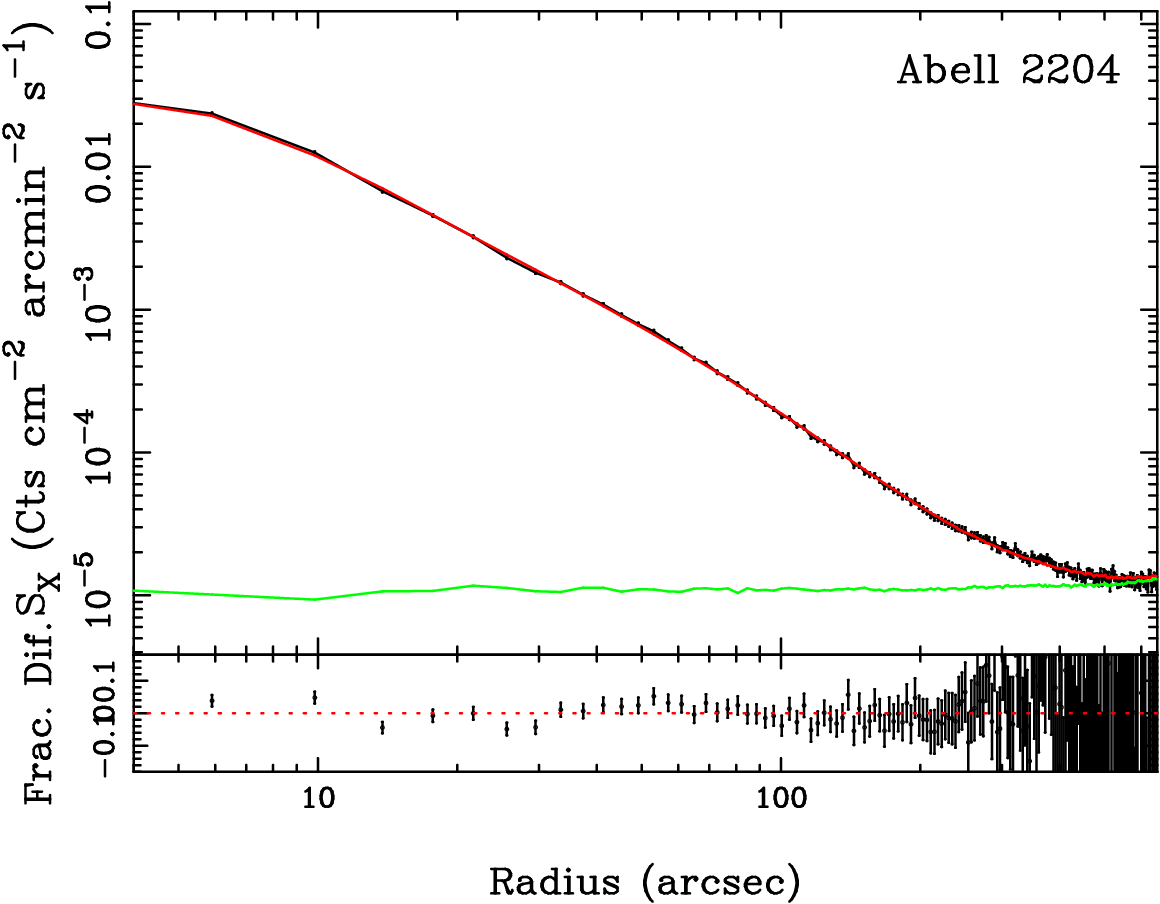}
\includegraphics[angle=-90,width=3in]{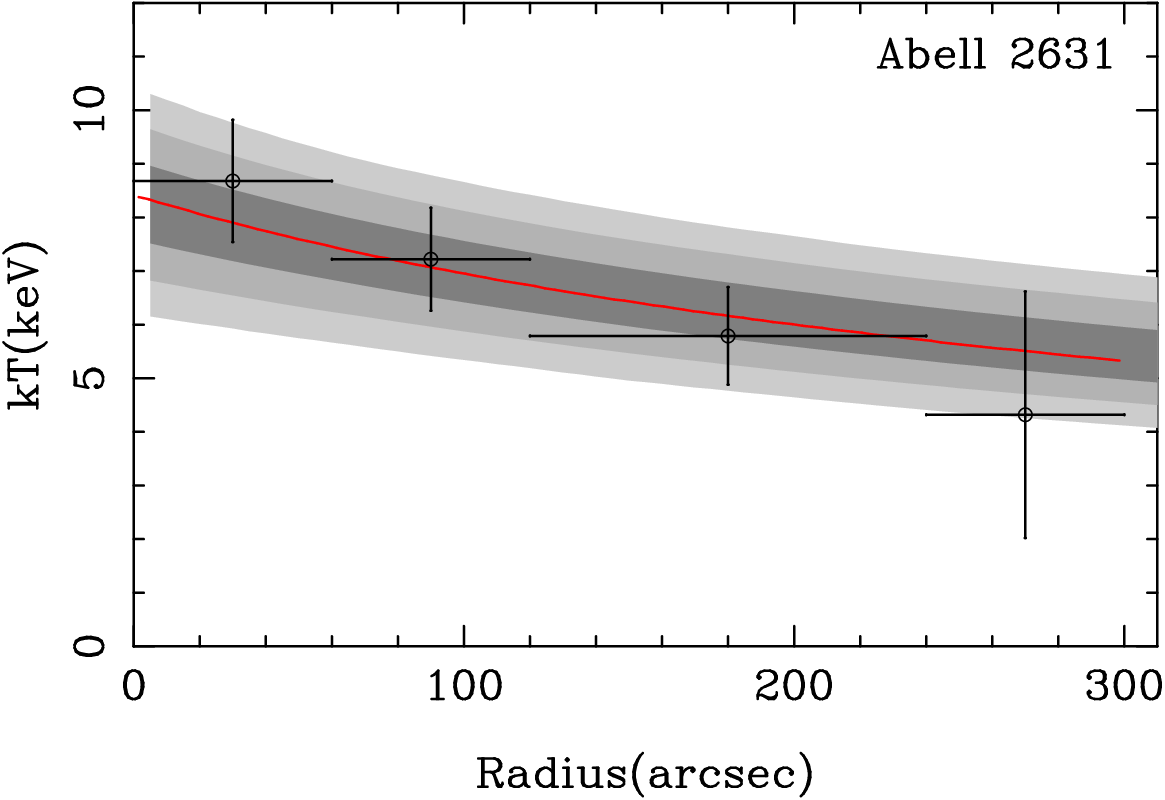}
\includegraphics[angle=-90,width=3in]{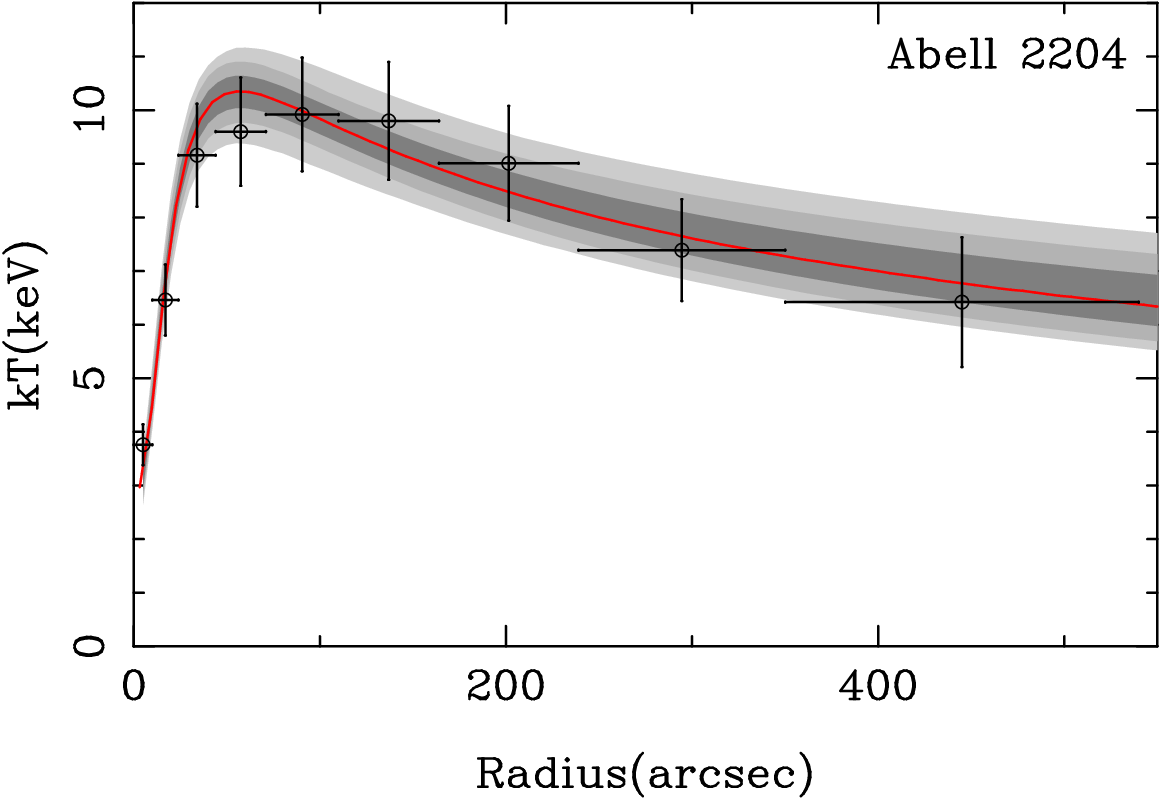}

\caption{
X-ray surface brightness and temperature profiles for Abell 2631 ({\it left column}) and Abell 2204 ({\it right column}). 
{\it Top panels:} Surface brightness profiles 
where the black points are derived from the X-ray images, the
red line shows the best fit model, and the green line is the background level determined from the blank sky observations.
The residuals show the fractional difference between the model and the data. A $1\%$ systematic
uncertainty has been added in quadrature to the datapoints.  We plotted the surface brightness profiles
beyond the fitted region to show the agreement between the cluster emission and background.
{\it Bottom panels:} Temperature profiles where the red line shows the best fit model and the dark, medium, and light grey regions show the 68\%, 95.4\%, and 99.7\% confidence levels obtained from the model fits.
A $10\%$ systematic uncertainty has been added in quadrature to the temperature bins.}
\label{fig:xray}
\vskip 12pt
\end{figure*}

\section{Modeling the X-ray and SZ data}
\label{sec:joint}

The X-ray observable is spatially resolved spectroscopy, and
temperature and metallicity of the intracluster plasma
 are measured using the X-ray spectroscopic data.
The X-ray  surface brightness is defined as
\begin{equation}
\Sx = \frac{1}{4 \pi (1+z)^3} \int n_{e}^{2} \Lamee(T_{e}, A) d\ell
\label{eqn:sx}
\end{equation}
where $\ell$ is the line of sight through the cluster, 
$n_{e}$ is the electron density, $T_e$ is the electron temperature, 
$A$ is the metallicity, and $\Lamee(T_e,A)$ is the X-ray cooling function 
(in units of counts cm$^3$ s$^{-1}$) as a function of 
electron temperature and metallicity.  The density,
temperature, and metallicity can vary along the line of sight.

The observable from the SZ data is the amplitude of the spectral 
distortion of the Cosmic Microwave Background (CMB) in the direction
of the cluster.  This distortion is due to inverse Compton scattering
of CMB photons off electrons in the intracluster medium (ICM), and 
results in a decrement in the 
CMB brightness temperature at frequencies $\lesssim~218$~GHz.  The magnitude 
of the decrement is proportional to the electron 
pressure integrated along the line of sight \citep{sunyaev1972}:
\begin{equation}
\Delta \Tsz = \Tcmb \int \sigT f(x,T_{e}) n_{e} \frac{k T_{e}}{m_e c^2}  d\ell
\label{eqn:t-sz}
\end{equation}
where \Tcmb\ is the temperature of the CMB, $f(x,T_{e})$ contains the
frequency dependence of the SZ temperature signature using the relativistic corrections provided by \cite{itoh1998} and \cite{nozawa2006}, 
 \sigT\ is the Thomson cross section,
$m_e$ is the electron mass, and $c$ is the speed of light.

Assuming spherical symmetry, the line of sight integration element 
$d \ell$ relates to the angular element (in radians) as $d\ell = \da d\theta$, where $\da$ is the angular diameter distance.
From Equations \ref{eqn:sx} and \ref{eqn:t-sz}, we find
\begin{eqnarray}
S_{x} \propto \int D_{A} n_{e}^2 \Lamee(T_{e}, A) d\theta \label{eq:propsx} \\
 \Delta \Tsz \propto \int D_{A} n_{e} T_{e} d\theta \label{eq:DeltaT}.
\end{eqnarray}
The combination of X-ray imaging spectroscopy and SZ observations
can be used to simultaneously measure
the distribution of the electron density, the electron
temperature, and the angular diameter distance
\citep[e.g.,][]{hughes1998,grego2000,reese2002,grainge2002,saunders2003,bonamente2006}. For further discussion on the SZ effect and its use for cosmology, see  reviews by \citet{birkinshaw1999} and
\citet{carlstrom2002}.

We describe the density and temperature profiles of the hot plasma in galaxy
clusters using the model proposed by \cite{bulbul2010}:
\begin{equation}
n_{e}(r)=n_{e0} \phi(r,r_{s},\beta)^n 
\tau_{cool}^{-1}
\label{eq:density}
\end{equation}
\begin{equation}
T(r)=T_{0} \phi(r,r_{s},\beta) 
\tau_{cool}
\label{eq:temp}
\end{equation}
where 
\begin{eqnarray}
\label{eqn:phi}
\phi(r,r_{s},\beta)=\frac{1}{(\beta-2)}\frac{(1+r/r_{s})^{\beta-2}-1}{r/r_{s}(1+r/r_{s})^{\beta-2}}, \\
\label{eqn:tcool}
\tau_{cool}=\frac{\alpha + (r/r_{\rm cool})^{\gamma}}{1 + (r/r_{\rm cool})^{\gamma}}, 
\end{eqnarray}
$n_{e0}$ is the normalization of the pressure profile, 
$n$ is the polytropic index, $r_{s}$ is the scale
radius, $T_{0}$ is the normalization factor for the scaling of the temperature profile,
{$\gamma$ is the slope of the cooling function, and $\alpha$ is the cooling parameter which ranges from 0 to 1.
One attractive feature of these models is that they provide a simple analytic
form for the electron pressure:
\begin{equation}
P_{e}(r)=P_{e0} \phi(r,r_{s},\beta)^{n+1},
\label{eqn:pres}
\end{equation}
where $P_{e0}=n_{e0} k  T_{0}$ is the pressure normalization.
The free parameters of the model, which are used to jointly
fit the X-ray and SZ data, are $n_{e0}, T_{0}, r_{s}, r_{\rm cool}, 
\alpha, \beta, \gamma, n$ and the distance $D_A$.

Parameter estimation is done using a Monte Carlo Markov chain (MCMC) method
described in \cite{bonamente2004}.
Correlation among model parameters is a feature of most analytic models,
including the one used in this paper. 
Parameter correlations can result in low acceptance rates and thus slow convergence
of the Markov chains \citep[e.g.,][]{gilks1996}.
In our implementation, steps of the MCMC are proposed along the set of
directions which diagonalize the covariance of the posterior distribuition,
determined via singular-value decomposition of an initial test chain,
resulting in efficient exploration of the parameter space (see Appendix).

\begin{figure}[!h]
\centering
\includegraphics[angle=-90,width=2.8in]{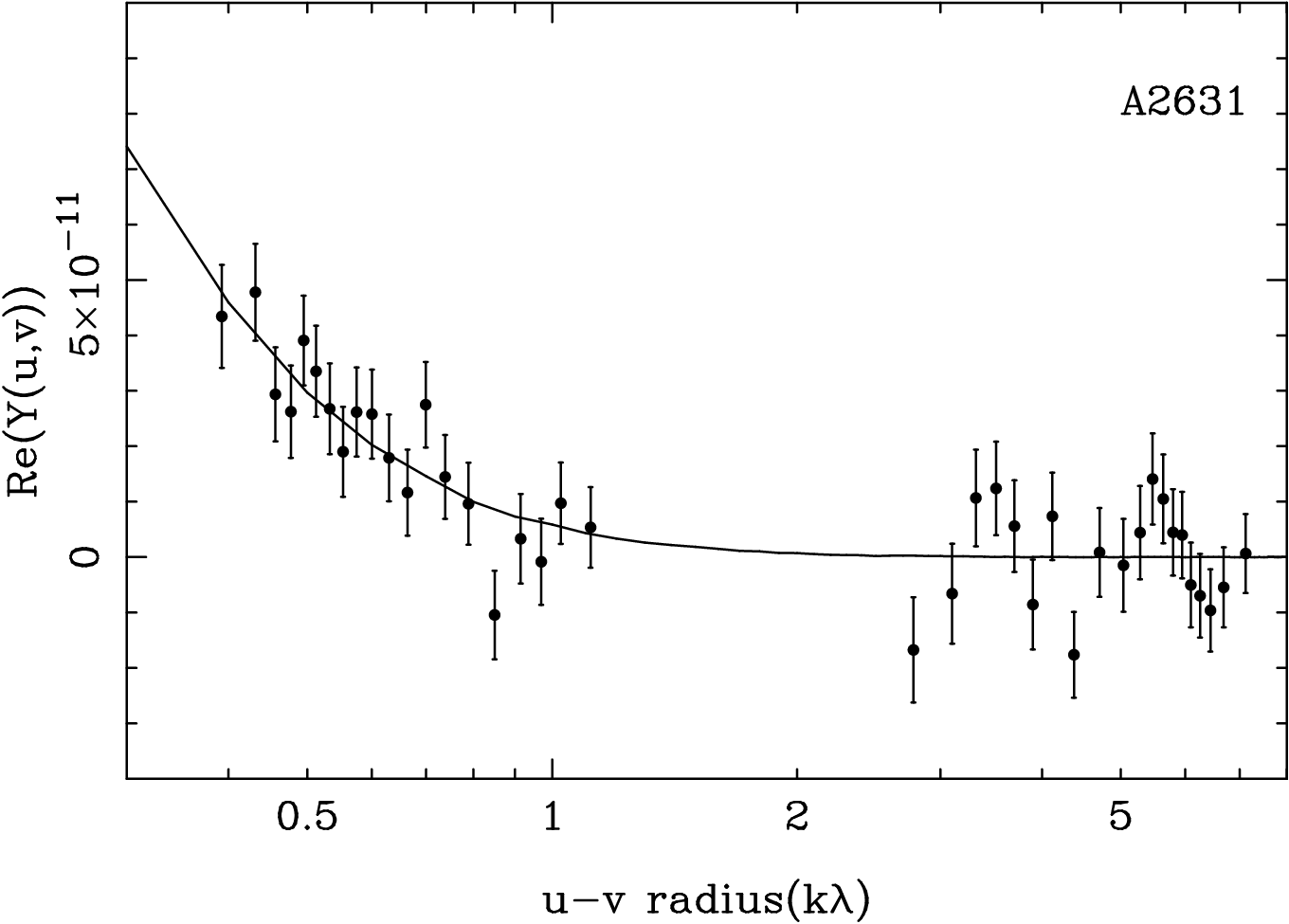}
\includegraphics[angle=-90,width=2.8in]{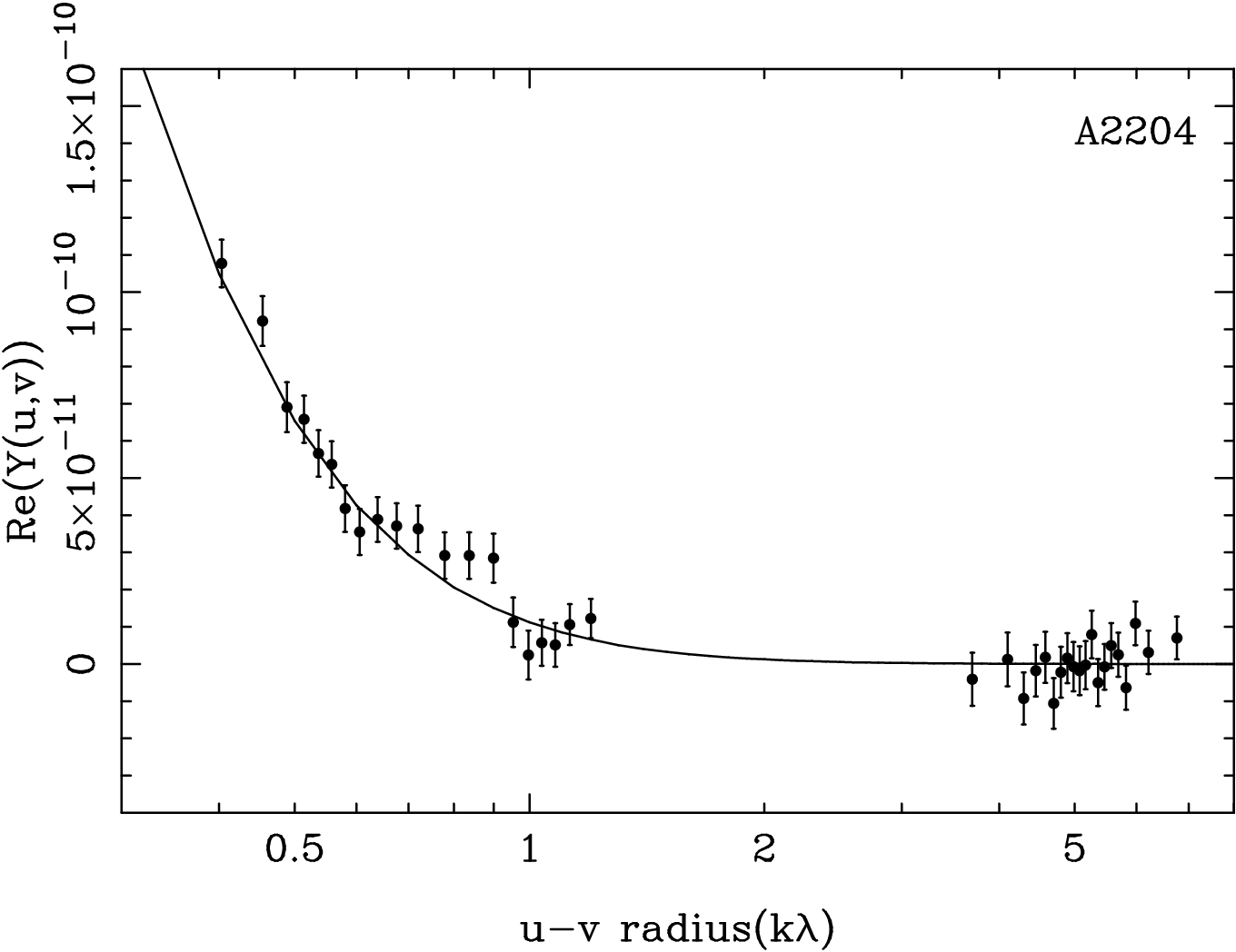}
\caption{SZ visibility profiles for \clusterA\ ({\it top}) and \clusterB\ ({\it bottom}) 
plotted as
a function of $u-v$ radius ($\sqrt{u^{2}+v^{2}}$).  The plots show the real components  of the measured
$Y(u,v)$  
along with the best fit model. }
\vskip 0.1 in
\label{fig:joint-sze}
\end{figure}

\subsection{X-ray Data Analysis}
\label{sec:analysis}
The annular bins in the temperature profile (Figure \ref{fig:xray}) 
were chosen by starting with
an initial 10$''$ bin and then increasing each bin by 50\% of the width of the previous 
bin to give roughly the same counts per bin.
Following the analysis of the systematic uncertainties for the X-ray data described
in \cite{bulbul2010}, we adopt a 1\% systematic uncertainty on the count rate
of each bin of the surface brightness profile and a 10\% systematic uncertainty on the temperature of each
spectral region as discussed in Section~\ref{sec:uncertainty-instru}.
Figure~\ref{fig:xray} shows the temperature and surface brightness profiles,
along with the best-fit models, for the fit to the X-ray data only of Abell~2631 and Abell~2204.
Since our model has the same parameters for both density and temperature,
the surface brightness profile carries a larger weight in the fit.
The model fits are acceptable for both clusters to within the plotted
errors, which include the systematic uncertainties associated with the surface brightness and temperature 
discussed in Section~\ref{sec:uncertainty}.
For Abell 2631, there is insufficient signal to constrain $n$ and $\beta$ 
simultaneously. In this case, we fix $\beta=2$, which is equivalent to assuming 
that the total mass follows a \cite{navarro1997} profile at large radii (see Section 4.3).

\subsection{SZ Data Analysis}
After removal of compact radio sources in the cluster field, 
the visibilities $V_\nu(u,v)$ measured by the SZA
can be related to the Fourier domain equivalent of the integrated Compton-$y$ 
parameter $Y(u,v)$ \citep[see, e.g.,][]{mroczkowski2009}.
This is defined as
\begin{equation}
\label{eq:Yuv}
Y(u,v) \equiv \frac{V_\nu(u,v)}{g(x) \, I_0},
\end{equation}
where $g(x)$ corrects for the frequency dependence of the SZ flux,
and $I_0 = 2 (k_B \Tcmb)^3/(h c)^2$ is the primary CMB intensity normalization.
Figure~\ref{fig:joint-sze} shows $Y(u,v)$ along with the best-fit model, for
the fit to the SZ data only of the two clusters.

\section{Joint Analysis of X-ray and SZ Data}
\label{sec:app}
We first perform
a consistency check of the determination of the pressure profiles from the 
X-ray and SZ data. We then focus on
determinations of the angular diameter distance
and the radial profile of the gas mass fraction using
the consistent parameterization of
density, temperature and pressure provided by the ICM model
for the joint analysis of the X-ray and SZ observables.
\pagebreak

\subsection{Consistency of X-ray and SZ measurements of the electron pressure profiles}
\label{sec:pressure}

X-ray and SZ observations provide independent
measurements of the radial distribution of the electron pressure. The X-ray observables
are the electron temperature and surface brightness, the latter depends on the square of the
electron density according to Equation~\ref{eqn:sx}; the SZ observable, on the other hand,
is directly proportional to the electron pressure integrated along the line of sight (Equation~\ref{eqn:t-sz}).

The two observables can be affected by different sources of systematic uncertainties.
For example,
the presence of non-thermal X-ray emission \citep[e.g.,][]{million2009,bonamente2005,sarazin1998}
could result in the increase of the X-ray emission above the level of the thermal gas,
and radio emission from cluster halos \citep[e.g.,][]{brunetti2007}
may partially fill the SZ decrement.
Another source of systematic uncertainty 
is the assumption of spherical symmetry  in the analysis \citep[e.g.,][]{sulkanen1999},
which would result in a different measurement of the pressure from X-ray and SZ observations.
A discussion of sources of systematic
uncertainty in the analysis of X-ray and SZ observations is presented in Section~\ref{sec:uncertainty}.
A comparison of the pressure profiles from SZ and X-ray observations is therefore useful to determine 
the presence of sources of emission that can cause differences between the
two measurements.

We perform a joint fit to the X-ray and SZ data
using Equation \ref{eqn:pres}, 
with the normalization of the 
SZ pressure model independent of the X-ray density and temperature 
normalizations. 
The common parameters in the models ($r_{s}$, $\beta$, and $n$) 
are linked between the two datasets, 
thus
requiring the X-ray and SZ pressure profiles to have the same shape.
In this analysis, we adopt the angular diameter 
distance appropriate for the cluster redshift in a $\Lambda$CDM model with $h=0.73$, $\Omega_M=0.27$ and 
$\Omega_{\Lambda}=0.73$.

The pressure inferred from the
X-ray and SZ measurements are within 20\% of one another for both clusters, 
consistent with the statistical and systematic effects (Tables \ref{tab:pres} and \ref{tab:unc});
results for a larger sample of clusters can be found in Bonamente et al. (2011).
Note that the measurement of the ratio of X-ray pressure to the SZ pressure depends on the
choice of the Hubble parameter, since the pressure normalizations are degenerate with the
value of $D_A$ assumed in the analysis.

\begin{figure*}
\centering
\includegraphics[angle=-90,width=2.8in]{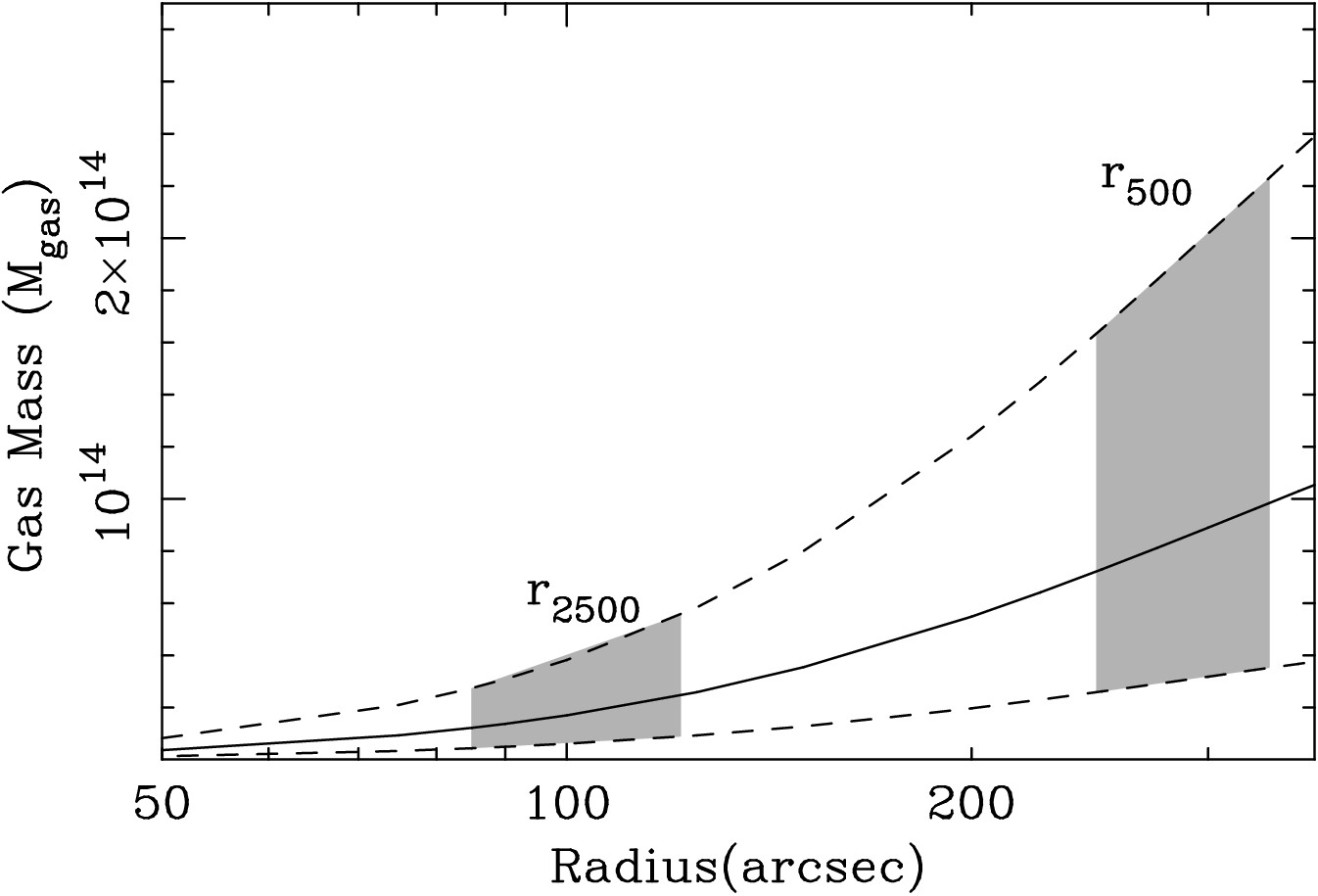}
\includegraphics[angle=-90,width=2.8in]{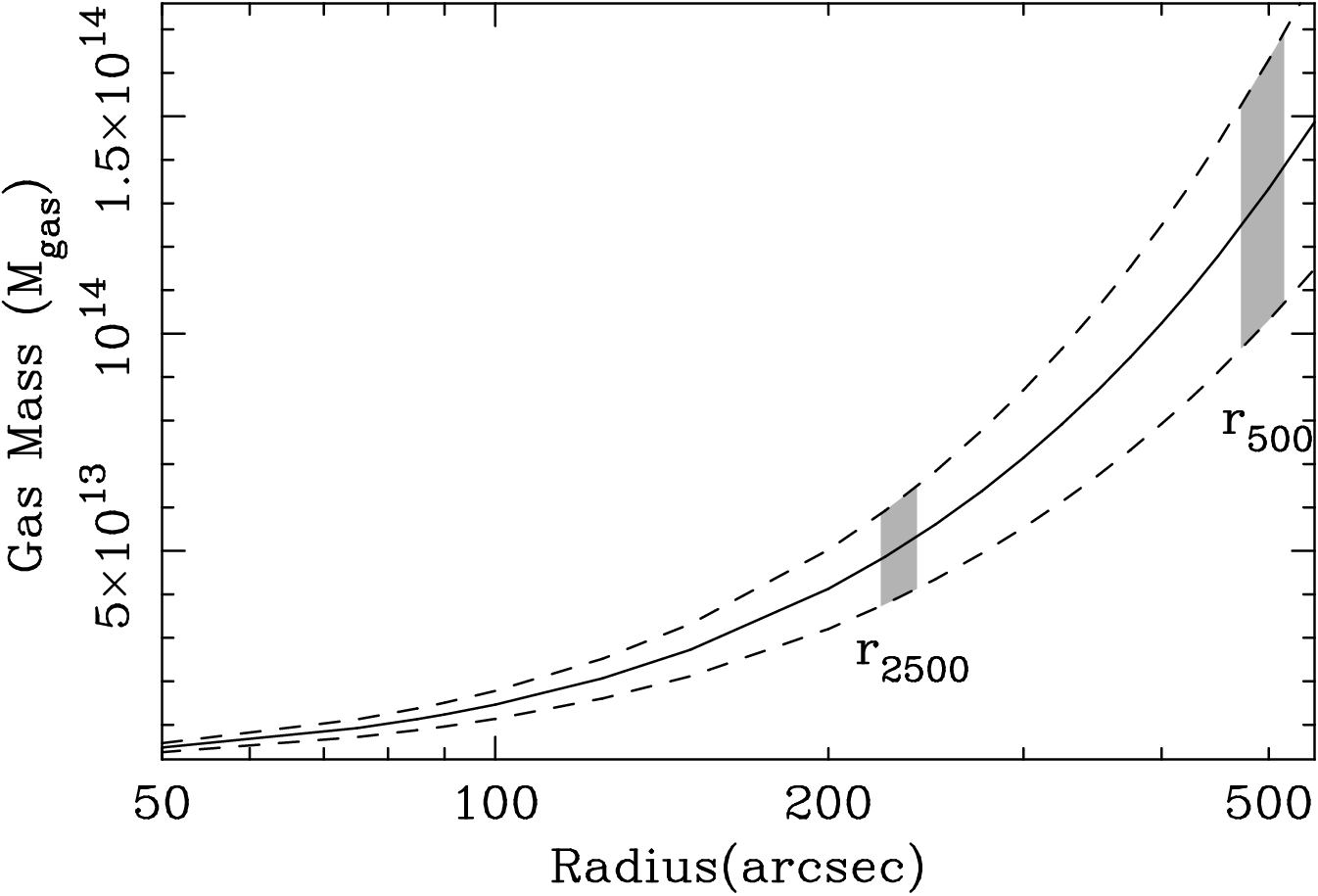}
\includegraphics[angle=-90,width=2.8in]{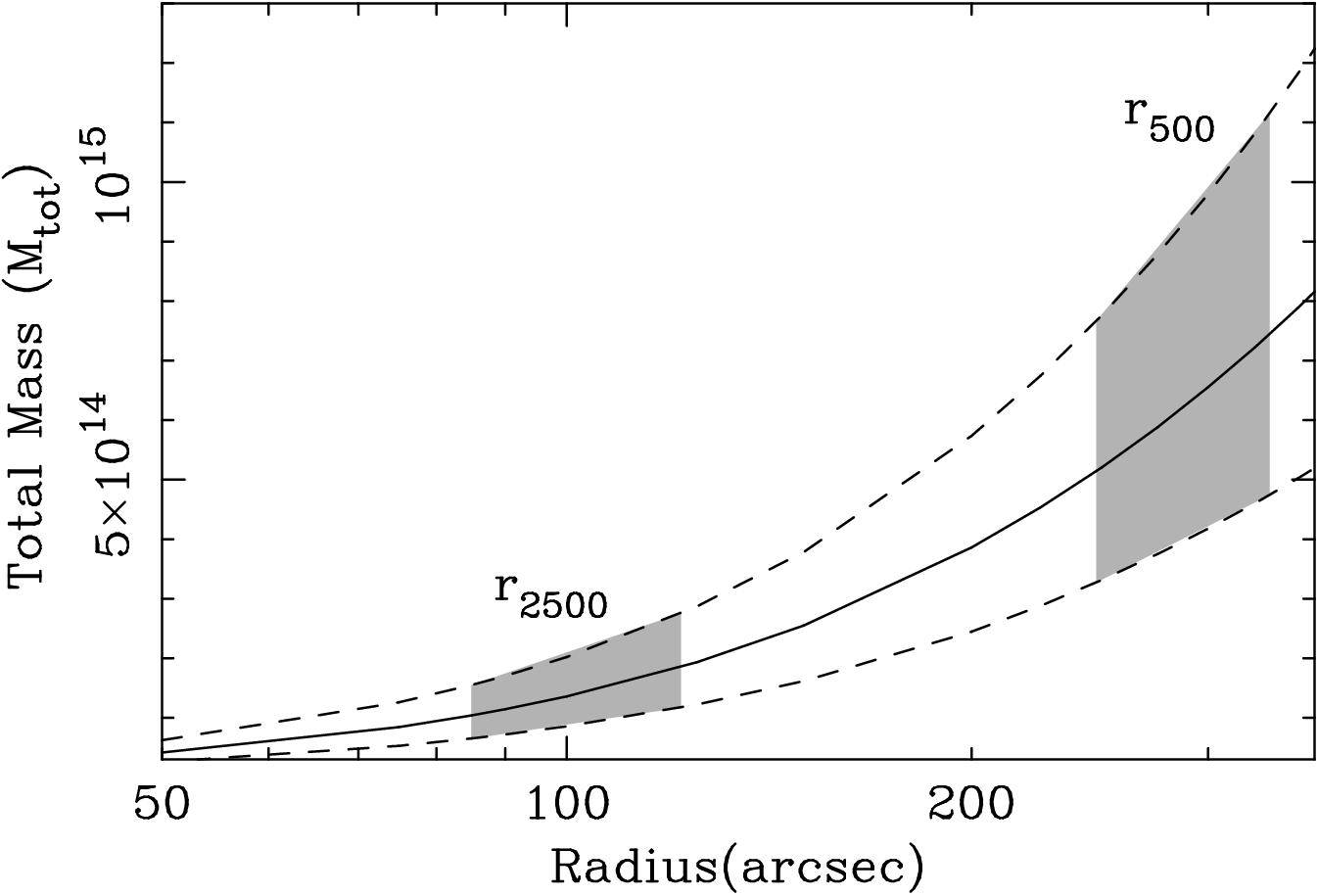}
\includegraphics[angle=-90,width=2.8in]{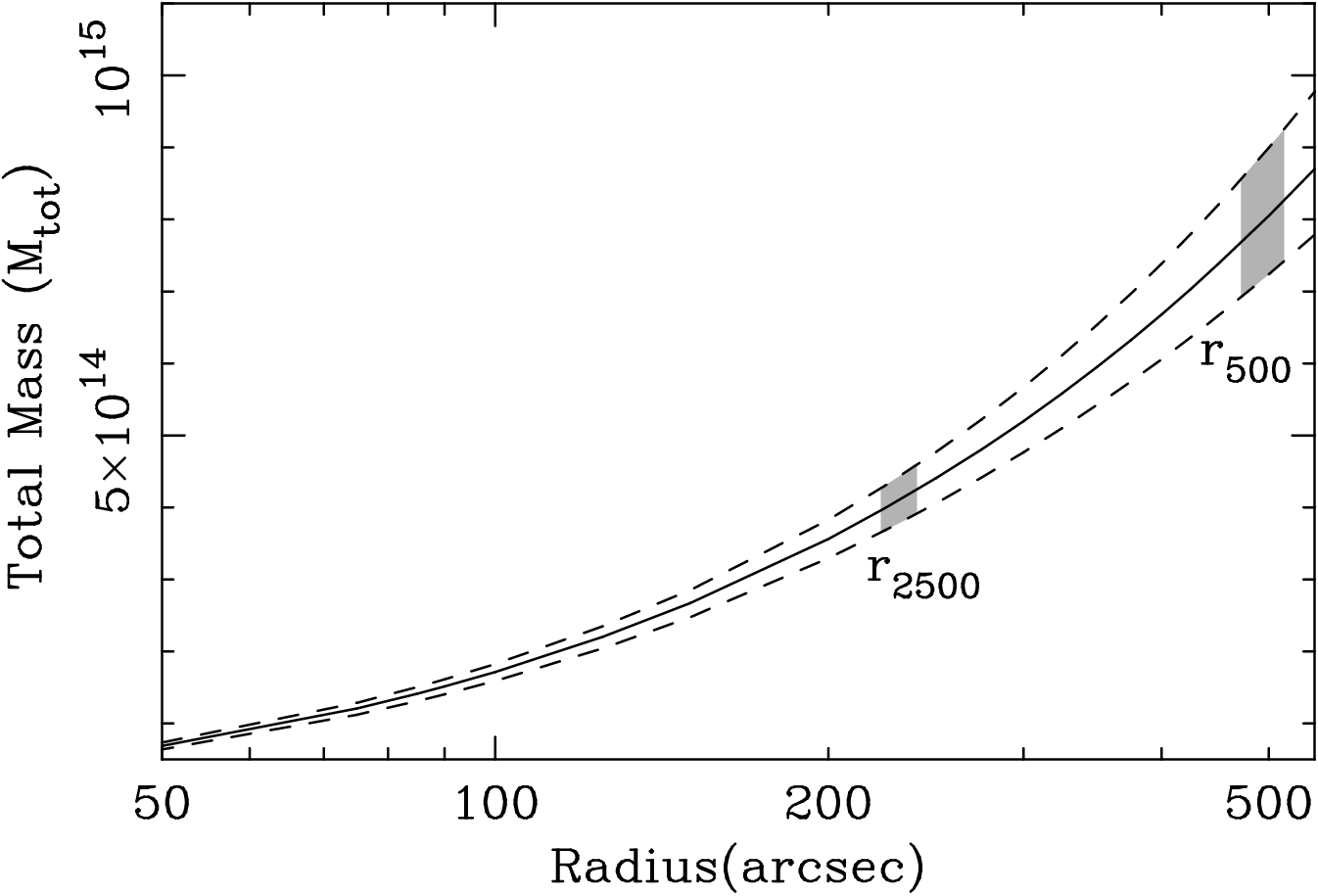}
\includegraphics[angle=-90,width=2.8in]{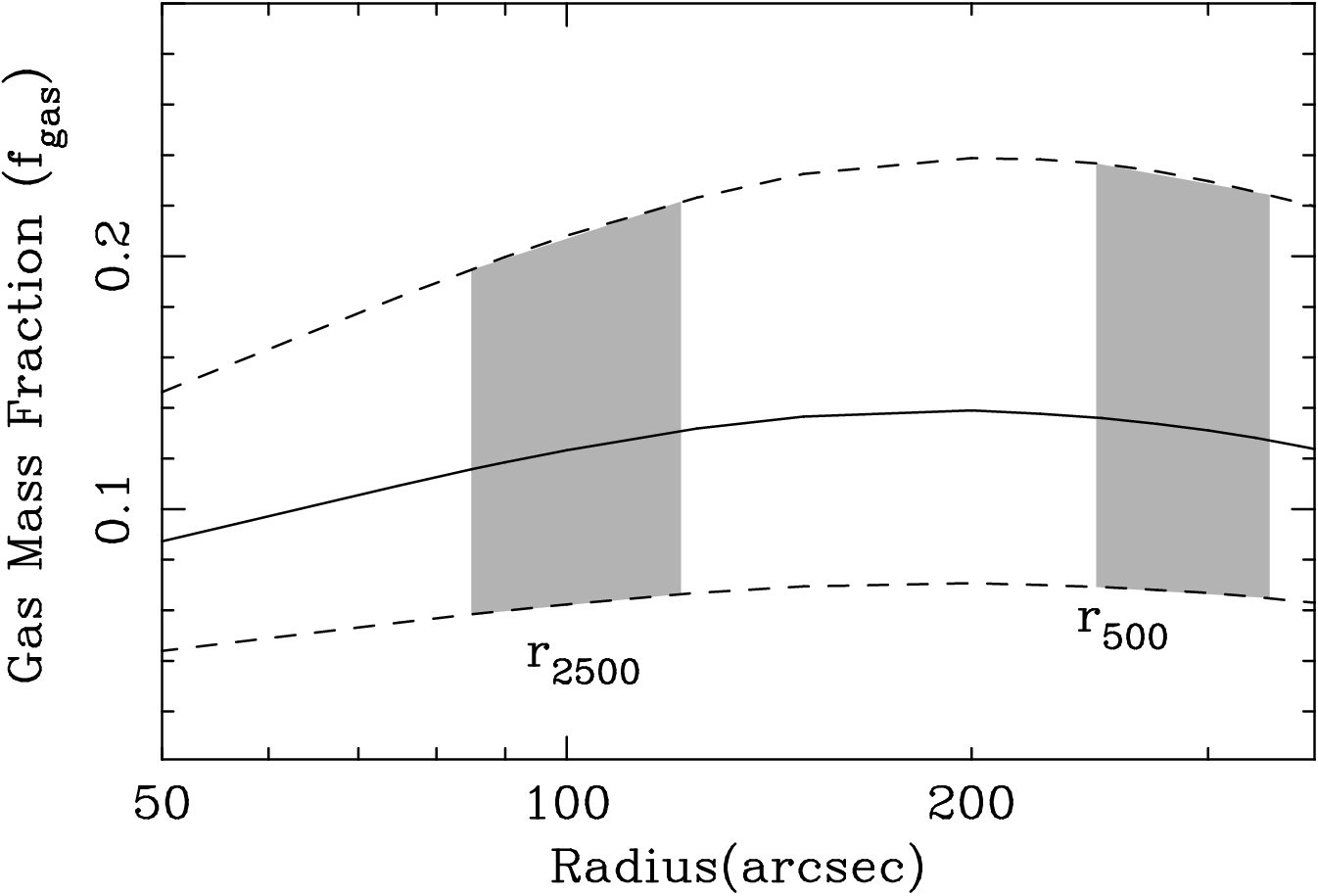}
\includegraphics[angle=-90,width=2.8in]{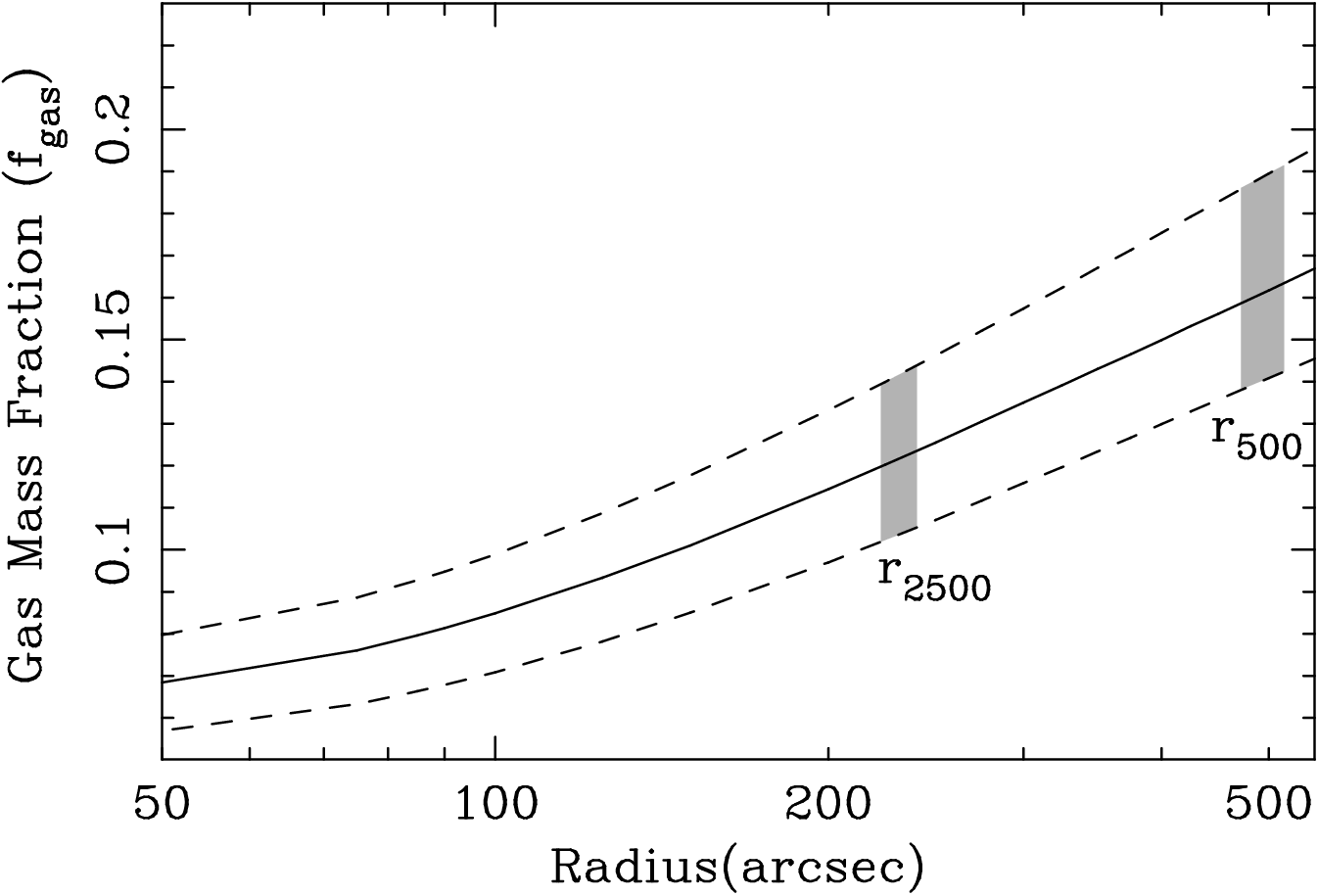}

\caption{\emph{Top Panels; left Abell~2631, right Abell~2204}: Gas mass profiles 
determined from the joint analysis of \chandra/SZA observations for Abell 2631 and
Abell 2204. 
\emph{Middle Panels}: Total mass
profiles.  
\emph{Bottom Panels}: Gas mass fraction profiles.  
The dashed lines are the 68\% confidence level at each radius. Grey areas show the measurements at
radii $r_{2500}$ and $r_{500}$, obtained
by marginalization over the cosmological parameters.  
\chandra\ calibration systematics are included in the measurements.
 }
\vspace{1.0cm}
\label{fig:fgas-joint}
\end{figure*}

\subsection{Direct measurement of the angular diameter distance}
\label{sec:da}

We also perform a joint X-ray and SZ analysis which enables us to place 
direct constraints on
the angular diameter distance without  using priors on the
cosmological parameters \citep[e.g.,][]{hughes1998,birkinshaw1999,reese2002,bonamente2006}.
For this analysis, we link the shape parameters and the pressure normalizations between the
X-ray and SZ data, and allow $D_{A}$ to vary.
For Abell 2631 and Abell 2204 
we measure $D_A=798.9\pm^{308.1}_{267.4} \rm{Mpc}$ and $D_{A}=575.3\pm^{46.6}_{55.6} \rm{Mpc}$,
both values are consistent with those calculated using a standard a $\Lambda$CDM cosmology at the 1--$\sigma$ level.
The measurement of the $D_A$ for Abell~2204 
is also in agreement with that of \cite{bonamente2006}.
The measurement of the angular diameter distance for a given cluster is affected by a
number of systematic effects \citep{bonamente2006}, and the agreement of $D_{A}$ with the $\Lambda$CDM
value is expected for a large sample but not necessarily for individual clusters, as it
is for the two clusters in this paper. 

\subsection{Radial profiles of the gas mass fraction independent of cosmological parameters}
\label{sec:fgas}
By using direct measurement of $D_A$ as described in Section~\ref{sec:da}, we can also 
obtain radial profiles of the gas mass, total mass, and the gas mass fraction without the need to use priors on the
cosmological parameters.
The gas mass \Mgas\ is computed by integrating the gas density profile within the volume,
\begin{equation}
\Mgas = 4 \pi \mu_e m_{p} \int n_{e}(r) r^{2} dr = 4 \pi \mu_e m_{p} \da\!^3 \int n_{e}(\theta) \theta^{2} d\theta,
\label{eqn:mgas}
\end{equation}
where $\mu$ is the mean molecular weight (calculated assuming metal abundances of $0.3$ 
solar, \citealt{anders1989}), $m_{p}$ is the proton mass, and $dr=\da d\theta$.
The total mass \Mtot\ is computed assuming hydrostatic equilibrium between the gravitational
mass and the thermal pressure of the gas:
\begin{equation}
\Mtot(r)= \frac{4\pi\rho_{i}r_{s}^{3}}{(\beta-2)}\left( \frac{1}{\beta-1} +\frac{1/(1-\beta) - r/r_s}{(1+r/r_{s})^{\beta-1}}\right)\tau_{c
ool}(r),
\label{eqn:mtot}
\end{equation}
where $\rho_{i}=(T_{0} k (n+1) (\beta-1))/(4 \pi G \mu m_{p} r_{s}^{2})$.  

Figure~\ref{fig:fgas-joint} shows
the radial profiles of the gas mass, total mass, and gas mass fraction for \clusterA\ and \clusterB.
The uncertainties reflect the fact that $D_A$ is also measured directly from the data,
and that no assumption about the value of the cosmological parameters $H_0$, $\Omega_M$ or 
$\Omega_{\Lambda}$ was made.

\subsection{Measurement of the gas mass fraction at an overdensity radius}
\label{sec:fgas-rdelta}
In cosmological applications 
\citep[e.g., via the distribution of \fgas\ with redshift,][]{allen2008}
\fgas\ is typically measured  within an overdensity radius.
The radius $r_{\Delta}$ is defined
as the radius within which the average matter density of the cluster
is $\Delta$ times the critical density of the universe at the cluster's redshift:
\begin{equation}
r_{\Delta}^{3} \equiv \frac{\Mtot(r_{\Delta})}{\frac{4\pi}{3} \Delta \rho_{c}(z)},
\label{eq:rdelta}
\end{equation}
where $\rho_{c}(z)=(3 H_0^2 E^2(z))/(8 \pi G)$
is the critical density of the universe, \ho\ is the Hubble constant,
and $E^2(z)$= $\OmegaM(1+z)^3 + \OmegaL + \Omegak(1+z)^2$
in the $\Lambda$CDM model.

The joint X-ray and SZ analysis provides
cosmology-independent constraints on $D_{A}$
and on the radial profile of $\fgas(r)$ (Sections~\ref{sec:da} and \ref{sec:fgas}).
The radius $r_{\Delta}$ and therefore all quantities calculated out to this radius
retain a cosmological dependence through the factor 
$\rho_{c}(z)$ appearing in Equation \ref{eq:rdelta}. 
In the following we describe a method to marginalize the measurement
of \fgas($r_{\Delta}$)  over the cosmological parameters, which results in 
a weak cosmology
dependence of our joint measurements of  \fgas($r_{\Delta}$).

In what follows, we adopt the standard
flat Friedman-Robertson-Walker (FRW)
cosmological model, parameterized by $H_{0}$ and \OmegaM, with $\OmegaL=1-\OmegaM$, 
which is  known to provide a good fit to other cosmic distance data \citep{allen2008,freedman2009,percival2010}.
Within this model, a given 
 $D_{A}$ value corresponds to a
curve in the ($H_{0}$, \OmegaM) plane described by
\begin{equation}
\label{eq:ho}
 H_0  = 
  \frac{c}{D_A(1+z)}\int_0^z \frac{d\zeta}{[\Omega_{M}((1+\zeta)^3-1)-1 ] ^{1/2} }, 
\end{equation}
where spatial
flatness is assumed. 
For each step in the Markov chain of each cluster,
we use the corresponding value of $D_A$ to generate a consistent pair of cosmological parameters by drawing a random value of 
\OmegaM\ from a uniform distribution on [0,1],
and calculating the associated value of $H_{0}$ from Equation~\ref{eq:ho}.
We then use these parameters when calculating $r_{\Delta}$ 
and the other fit parameters of the chain. This approach
has the advantage of using minimal prior information on the
cosmological parameters, with \OmegaM\ free to range between 0 and 1,
and $D_A$ measured directly from the data.
This method properly
accounts for the covariance between $r_{\Delta}$, $\fgas(r_{\Delta})$, $D_{A}$ and the
other parameters of the fit.
For the flat FRW model and the
low redshifts in question, this procedure corresponds, to very good
approximation, to the use of uniform priors on $H_{0}$ and \OmegaM. We note,
however, that this correspondence does not necessarily hold for higher redshifts
($z\sim$1) or for other cosmological models.
The measurements of masses and $\fgas$ 
at $r_{2500}$ and $r_{500}$ are indicated as grey areas in Figure \ref{fig:fgas-joint} and are listed in
Table \ref{tab:joint}.

For comparison, we also provide the measurements of masses and \fgas\ using
the X-ray data with
a fixed $D_A$ calculated from the standard flat $\Lambda$CDM cosmology, and
priors on $\Omega_M$ from the WMAP measurements of \cite{komatsu2011}.
The results are reported in Table~\ref{tab:xray}.

\subsection{Sources of Systematic Uncertainty} 
\label{sec:uncertainty}
The sources of systematic uncertainty on the gas mass fraction and the
integrated X-ray and SZ pressure are listed in Table \ref{tab:unc}.  The
individual errors are added in quadrature to determine the total
systematic uncertainty on $f_{gas}$ and the integrated pressure.
Systematics from \chandra\ instrument calibration and the X-ray background 
are included in the fitting of the data, and the value of masses and $f_{gas}$ in Tables
\ref{tab:joint} and \ref{tab:xray} account for these systematics.
Since this joint analysis method can be applied to larger samples of clusters, 
we also indicate whether the impact of each source of systematic uncertainty 
is reduced with sample size.

\subsubsection{Instrument Calibration} 
\label{sec:uncertainty-instru}
\chandra's ACIS effective area has a spatially dependent non-uniformity at the level 
of $\pm1\%$ and therefore we add a $\pm1\%$ uncertainty to the surface brightness data.
We also adopt a $\pm10\%$ uncertainty on the temperature measurements to account for uncertainty
in the low-energy calibration of the effective area
\citep[see, e.g.,][]{bulbul2010}.   
These uncertainty estimates for the \chandra\ calibration are folded into the mass measurements  
reported in Tables \ref{tab:pres}, \ref{tab:joint} and \ref{tab:xray}. 

Frequent observations of Mars are used to calibrate the SZA absolute flux scale; we employ
the \cite{rudy1987} flux model which has an estimated absolute calibration uncertainty of
$\pm5\%$.  The stability of the instrumental gain is $\pm3\%$, as determined from repeated
calibrator measurements in SZA survey fields \citep{muchovej2011}.
The absolute calibration and instrumental gain yield a global $\pm6\%$ uncertainty on the
SZA calibration.  We rescaled the SZA data by the $\pm6\%$ uncertainty on the SZA calibration and compared the measurements to the 
original analysis.  We found a $\pm6\%$ systematic uncertainty on the pressure and $\pm10\%$ systematic uncertainty on the gas mass fraction at $r_{2500}$ and $r_{500}$.
The uncertainty associated with instrument calibration does not average down 
with sample size.

\subsubsection{Kinetic SZ Effect}
\cite{reese2002} reports that for a cluster with a temperature of 8.0 keV
and with a typical velocity along the line of sight of 300 km/s \citep{watkins1997,colberg2000}
the kinetic SZ effect would be $\pm4\%$ of the thermal SZ for 30 GHz observations.
Accordingly we use a $\pm4\%$ uncertainty due to the kinetic SZ effect on the gas mass
fraction and the SZ pressure profiles for measurements at $r_{2500}$ and $r_{500}$.
This 
source of uncertainty averages down by a factor of 
the square root of the size of the sample.

\subsubsection{Radio Source Contamination}
Undetected radio sources not accounted in the modeling could lead to a biased 
measurement of the SZ decrement.  We use 
the Faint Images of the Radio Sky at Twenty-centimeters (FIRST)
database as a reference for locating compact radio sources within 10$'$ of the cluster center.
Most radio sources that will affect the 30 GHz data (rms noise $\sim$ 0.25 mJy) will have counterparts in the
FIRST survey (rms noise of 0.15 mJy at 1.4 GHz);  inverted spectrum sources may not have
counterparts at 1.4 GHz and will affect our measurement of the SZ decrement, 
but they comprise a small fraction of the source population \citep[][Fig. 3]{muchovej2010}.

We determine the effect of undetected point sources by 
placing a radio source model at each FIRST source, fixing the position and marginalizing 
over the flux.  In the pressure model (Equation 7), we fixed the parameters $r_{s}$, $n$, $\beta$, and $D_{A}$ and let
$P_{eo}$ be free.  We compare the pressure profiles with the original analysis (see Table 2) 
and find a $\le 1\%$ difference in the pressure of each cluster.  Therefore we conservatively 
apply a $1\%$ uncertainty on the
pressure and $2\%$ uncertainty on the gas mass fraction at $r_{2500}$ and $r_{500}$.

\subsubsection{Asphericity}
\label{sec:systematics-asphericity}
Although we assume a spherical model in our analysis, most clusters do not appear to be
circular in shape in X-ray or radio observations. 
\cite{laroque2006} reports a $10-20\%$ uncertainty in the measurement of the gas mass fraction due to
asphericity; therefore we use $\pm20\%$ as a conservative estimate for measurements at $r_{2500}$ and $r_{500}$.
This uncertainty also averages down by a factor of the square root of the sample size, as shown in \cite{sulkanen1999},
provided the selection of the sample is unbiased with respect to cluster shape.

\subsubsection{Hydrostatic Equilibrium Assumption}
The assumption of hydrostatic equilibrium at large radii results in an
underestimate of the total mass.  This is due to the presence of non-thermal
pressure which can bias hydrostatic equilibrium measurements of the total mass (Evrard et al. 1990).
According to \cite{lau2009} the total mass of a relaxed cluster such 
as Abell~2204 will be biased by $-6\%$ at $r_{2500}$ and 
 $-8\%$ at $r_{500}$, and  for unrelaxed systems such as Abell~2631 
by $-9\%$ at $r_{2500}$ and $-11\%$ at $r_{500}$.
Therefore we adopt a 
systematic uncertainty of $-9\%$ and $-11\%$ in our error analysis for
measurements at $r_{2500}$ and $r_{500}$.

\subsubsection{X-ray Background}
The X-ray background is determined from the ACIS blank sky composite event file.  We normalize
the blank sky background level for each observation using an emission-free region on the
ACIS detector.  
We adjusted the background normalization factor by a factor of $\pm2\sigma$ and propagated this through
the analysis, and found that this produces a $\pm2\%$ uncertainty on the background count rate.  This uncertainty
affects the surface brightness and temperature measurements resulting in a $\pm2\%$ and $\pm9\%$ uncertainty on
the gas mass fraction measurements at $r_{2500}$ and $r_{500}$, and a $\pm2\%$ uncertainty on the X-ray pressure profiles at both radii.
This uncertainty averages down by a factor of the square root of the sample size.

\subsubsection{Systematics associated with the use of the \cite{bulbul2010} models}
The \cite{bulbul2010} model assumes a polytropic relationship
between the ICM density and temperature at large radii.
To estimate  uncertainties associated with the polytropic
assumption, we
compare our X-ray masses for the two clusters in our sample with those calculated using the \cite{vikhlinin2006}
model, which provides an independent parameterization of the
thermodynamic quantities.
From this comparison, we find that the gas mass fraction measurements varies by $\leq 10\%$ at all radii
between the 
\cite{bulbul2010} and \cite{vikhlinin2006} models. 
We estimate the uncertainty on the corresponding 
pressure profiles by comparing the integrated pressures between the two models, 
and the comparison results in  an uncertainty of $\pm3\%$ at $r_{2500}$ and $\pm5\%$ at $r_{500}$.
We consider these uncertainties as a rough estimate of the systematics associated with the 
\cite{bulbul2010} model.

\subsubsection{Helium Sedimentation}
The effect of helium sedimentation may be an additional
source of systematic uncertainty.  In our measurements 
we assume that the hydrogen to helium ratio is uniform
throughout the cluster.  However, theoretical studies
\citep{fabian1977,rephaeli1978} suggest
helium sedimentation effects may affect cluster mass measurements.  
\cite{peng2009} finds that the
bias in gas mass fraction from the presence of helium sedimentation
is less than 10\% at $\sim r_{2500}$ and negligible at $r_{500}$.
Accordingly, we estimate a systematic uncertainty of 10\% at $r_{2500}$ and
$\leq$5\% at $r_{500}$ for the measurement of the gas mass fraction.
\cite{bulbul2011} applied
the \cite{peng2009}  helium sedimentation simulation model
to a sample of clusters and demonstrated the effects on the
gas mass and total mass. 
The integrated pressure is proportional to the gas mass, and we use the values from \cite{bulbul2011} 
to determine upper limits to the systematic uncertainty 
in the measurement of pressure of
 $-4\%$ at $r_{2500}$ and $-2\%$ at $r_{500}$.

\section{Conclusions}
\label{sec:disconc}

We demonstrate the use of the \cite{bulbul2010} cluster model for
simultaneous fitting of X-ray data and Sunyaev-Zel'dovich effect data.
The model employs a compact parameterization that relates the three
primary thermodynamic quantities by the ideal gas law at all radii. We
consider X-ray data from \chandra\ and 30-GHz SZ data from the SZA for
both clusters, \clusterA\ and \clusterB\
and find that the model
adequately captures the radial variation in both the X-ray surface
brightness and SZ Compton-$y$ profiles.  For all clusters, separate
determinations of the electron pressure from the X-ray and SZ data
yield profiles that are statistically consistent.

Joint analysis of the X-ray and SZ data provides a direct measure of
$D_A$, the angular diameter distance to the cluster, that is
independent of cosmology.  For both clusters, this analysis yields a
measure of $D_A$ that is consistent with the standard \LCDM\
values at the 1-$\sigma$ level.  Using the measured angular diameter distance as a
constraint between $H_0$ and \OmegaM, we marginalize over the implicit
cosmology dependence of the overdensity radius to obtain estimates
of \fgas\ at $r_{2500}$ and $r_{500}$ that are only weakly dependent
on \OmegaM.

We discuss possible sources of systematic errors in the \fgas\
determination, and find that most will be mitigated if \fgas\ is
averaged over a large sample of clusters.  
A sample spanning a large redshift range can be used to constrain the evolution of \fgas\ with
redshift, and for constraining cosmological models with clusters
\citep[e.g.,][]{sasaki1996,pen1997,reese2002,allen2004,laroque2006,bonamente2006,allen2008,ettori2009}.

\section*{Acknowledgments}
The operation of the SZA is
supported by NSF through grant
AST-0604982 and AST-0838187.  Partial support is also provided from grant PHY-0114422 at the University of Chicago, and by NSF grants AST-0507545
and AST-05-07161 to Columbia University.  CARMA operations are supported by the NSF
under a cooperative
agreement, and by the CARMA partner universities.  SM acknowledges support from an NSF
Astronomy and Astrophysics Fellowship; CG and SM  from NSF
Graduate Research Fellowships; DPM from NASA Hubble Fellowship grant HF-51259.01.
Support for this work was provided for TM by NASA through the Einstein Fellowship Program, grant PF0-110077.

\bibliographystyle{apj}

\begin{thebibliography}{60}
\expandafter\ifx\csname natexlab\endcsname\relax\def\natexlab#1{#1}\fi

\bibitem[{{Allen} {et~al.}(2008){Allen}, {Rapetti}, {Schmidt}, {Ebeling},
  {Morris}, \& {Fabian}}]{allen2008}
{Allen}, S.~W., {Rapetti}, D.~A., {Schmidt}, R.~W., {Ebeling}, H., {Morris},
  R.~G., \& {Fabian}, A.~C. 2008, \mnras, 383, 879

\bibitem[{{Allen} {et~al.}(2004){Allen}, {Schmidt}, {Ebeling}, {Fabian}, \&
  {van Speybroeck}}]{allen2004}
{Allen}, S.~W., {Schmidt}, R.~W., {Ebeling}, H., {Fabian}, A.~C., \& {van
  Speybroeck}, L. 2004, \mnras, 353, 457

\bibitem[{{Anders} \& {Grevesse}(1989)}]{anders1989}
{Anders}, E., \& {Grevesse}, N. 1989, \gca, 53, 197

\bibitem[{{Benson} {et~al.}(2011){Benson}, {de Haan}, {Dudley}, {Reichardt},
  {Aird}, {Andersson}, {Armstrong}, {Bautz}, {Bayliss}, {Bazin}, {Bleem},
  {Brodwin}, {Carlstrom}, {Chang}, {Cho}, {Clocchiatti}, {Crawford}, {Crites},
  {Desai}, {Dobbs}, {Foley}, {Forman}, {George}, {Gladders}, {Halverson},
  {High}, {Holder}, {Holzapfel}, {Hoover}, {Hrubes}, {Jones}, {Joy}, {Keisler},
  {Knox}, {Lee}, {Leitch}, {Liu}, {Lueker}, {Luong-Van}, {Mantz}, {Marrone},
  {McDonald}, {McMahon}, {Mehl}, {Meyer}, {Mocanu}, {Mohr}, {Montroy},
  {Murray}, {Natoli}, {Padin}, {Plagge}, {Pryke}, {Rest}, {Ruel}, {Ruhl},
  {Saliwanchik}, {Saro}, {Schaffer}, {Shaw}, {Shirokoff}, {Song}, {Spieler},
  {Stalder}, {Staniszewski}, {Stark}, {Story}, {Stubbs}, {Suhada}, {van
  Engelen}, {Vanderlinde}, {Vieira}, {Vikhlinin}, {Williamson}, {Zahn}, \&
  {Zenteno}}]{benson2011}
{Benson}, B.~A. {et~al.} 2011, ArXiv e-prints

\bibitem[{{Birkinshaw}(1999)}]{birkinshaw1999}
{Birkinshaw}, M. 1999, Physics Reports, 310, 97

\bibitem[{{Bonamente} {et~al.}(2004){Bonamente}, {Joy}, {Carlstrom}, {Reese},
  \& {LaRoque}}]{bonamente2004}
{Bonamente}, M., {Joy}, M.~K., {Carlstrom}, J.~E., {Reese}, E.~D., \&
  {LaRoque}, S.~J. 2004, \apj, 614, 194

\bibitem[{{Bonamente} {et~al.}(2006){Bonamente}, {Joy}, {LaRoque}, {Carlstrom},
  {Reese}, \& {Dawson}}]{bonamente2006}
{Bonamente}, M., {Joy}, M.~K., {LaRoque}, S.~J., {Carlstrom}, J.~E., {Reese},
  E.~D., \& {Dawson}, K.~S. 2006, \apj, 647, 25

\bibitem[{{Bonamente} {et~al.}(2005){Bonamente}, {Lieu}, {Mittaz}, {Kaastra},
  \& {Nevalainen}}]{bonamente2005}
{Bonamente}, M., {Lieu}, R., {Mittaz}, J.~P.~D., {Kaastra}, J.~S., \&
  {Nevalainen}, J. 2005, \apj, 629, 192

\bibitem[{{Brunetti} {et~al.}(2007){Brunetti}, {Venturi}, {Dallacasa},
  {Cassano}, {Dolag}, {Giacintucci}, \& {Setti}}]{brunetti2007}
{Brunetti}, G., {Venturi}, T., {Dallacasa}, D., {Cassano}, R., {Dolag}, K.,
  {Giacintucci}, S., \& {Setti}, G. 2007, \apjl, 670, L5

\bibitem[{{Bulbul} {et~al.}(2010){Bulbul}, {Hasler}, {Bonamente}, \&
  {Joy}}]{bulbul2010}
{Bulbul}, G.~E., {Hasler}, N., {Bonamente}, M., \& {Joy}, M. 2010, \apj, 720,
  1038

\bibitem[{{Bulbul} {et~al.}(2011){Bulbul}, {Hasler}, {Bonamente}, {Joy},
  {Marrone}, {Miller}, \& {Mroczkowski}}]{bulbul2011}
{Bulbul}, G.~E., {Hasler}, N., {Bonamente}, M., {Joy}, M., {Marrone}, D.,
  {Miller}, A., \& {Mroczkowski}, T. 2011, ArXiv e-prints

\bibitem[{{Carlstrom} {et~al.}(2002){Carlstrom}, {Holder}, \&
  {Reese}}]{carlstrom2002}
{Carlstrom}, J.~E., {Holder}, G.~P., \& {Reese}, E.~D. 2002, \araa, 40, 643

\bibitem[{{Cavaliere} \& {Fusco-Femiano}(1976)}]{cavaliere1976}
{Cavaliere}, A., \& {Fusco-Femiano}, R. 1976, \aap, 49, 137

\bibitem[{{Colberg} {et~al.}(2000){Colberg}, {White}, {MacFarland}, {Jenkins},
  {Pearce}, {Frenk}, {Thomas}, \& {Couchman}}]{colberg2000}
{Colberg}, J.~M., {White}, S.~D.~M., {MacFarland}, T.~J., {Jenkins}, A.,
  {Pearce}, F.~R., {Frenk}, C.~S., {Thomas}, P.~A., \& {Couchman}, H.~M.~P.
  2000, \mnras, 313, 229

\bibitem[{{Ettori} {et~al.}(2009){Ettori}, {Morandi}, {Tozzi}, {Balestra},
  {Borgani}, {Rosati}, {Lovisari}, \& {Terenziani}}]{ettori2009}
{Ettori}, S., {Morandi}, A., {Tozzi}, P., {Balestra}, I., {Borgani}, S.,
  {Rosati}, P., {Lovisari}, L., \& {Terenziani}, F. 2009, \aap, 501, 61

\bibitem[{{Fabian} \& {Pringle}(1977)}]{fabian1977}
{Fabian}, A.~C., \& {Pringle}, J.~E. 1977, \mnras, 181, 5P

\bibitem[{{Freedman} {et~al.}(2009){Freedman}, {Burns}, {Phillips}, {Wyatt},
  {Persson}, {Madore}, {Contreras}, {Folatelli}, {Gonzalez}, {Hamuy}, {Hsiao},
  {Kelson}, {Morrell}, {Murphy}, {Roth}, {Stritzinger}, {Sturch}, {Suntzeff},
  {Astier}, {Balland}, {Bassett}, {Boldt}, {Carlberg}, {Conley}, {Frieman},
  {Garnavich}, {Guy}, {Hardin}, {Howell}, {Kessler}, {Lampeitl}, {Marriner},
  {Pain}, {Perrett}, {Regnault}, {Riess}, {Sako}, {Schneider}, {Sullivan}, \&
  {Wood-Vasey}}]{freedman2009}
{Freedman}, W.~L. {et~al.} 2009, \apj, 704, 1036

\bibitem[{{Gilks} {et~al.}(1996){Gilks}, {Richardson}, \&
  {Spiegelhalter}}]{gilks1996}
{Gilks}, W., {Richardson}, S., \& {Spiegelhalter}, D. 1996, Markov Chain Monte
  Carlo in Practice (Chapman and Hall)

\bibitem[{{Grainge} {et~al.}(2002){Grainge}, {Grainger}, {Jones}, {Kneissl},
  {Pooley}, \& {Saunders}}]{grainge2002}
{Grainge}, K., {Grainger}, W.~F., {Jones}, M.~E., {Kneissl}, R., {Pooley},
  G.~G., \& {Saunders}, R. 2002, \mnras, 329, 890

\bibitem[{{Grego} {et~al.}(2000){Grego}, {Carlstrom}, {Joy}, {Reese}, {Holder},
  {Patel}, {Cooray}, \& {Holzapfel}}]{grego2000}
{Grego}, L., {Carlstrom}, J.~E., {Joy}, M.~K., {Reese}, E.~D., {Holder}, G.~P.,
  {Patel}, S., {Cooray}, A.~R., \& {Holzapfel}, W.~L. 2000, \apj, 539, 39

\bibitem[{{Hickox} \& {Markevitch}(2006)}]{hickox2006}
{Hickox}, R.~C., \& {Markevitch}, M. 2006, \apj, 645, 95

\bibitem[{{Hughes} \& {Birkinshaw}(1998)}]{hughes1998}
{Hughes}, J.~P., \& {Birkinshaw}, M. 1998, \apj, 501, 1

\bibitem[{{Itoh} {et~al.}(1998){Itoh}, {Kohyama}, \& {Nozawa}}]{itoh1998}
{Itoh}, N., {Kohyama}, Y., \& {Nozawa}, S. 1998, \apj, 502, 7

\bibitem[{{Komatsu} {et~al.}(2011){Komatsu}, {Smith}, {Dunkley}, {Bennett},
  {Gold}, {Hinshaw}, {Jarosik}, {Larson}, {Nolta}, {Page}, {Spergel},
  {Halpern}, {Hill}, {Kogut}, {Limon}, {Meyer}, {Odegard}, {Tucker}, {Weiland},
  {Wollack}, \& {Wright}}]{komatsu2011}
{Komatsu}, E. {et~al.} 2011, \apjs, 192, 18

\bibitem[{{LaRoque} {et~al.}(2006){LaRoque}, {Bonamente}, {Carlstrom}, {Joy},
  {Nagai}, {Reese}, \& {Dawson}}]{laroque2006}
{LaRoque}, S.~J., {Bonamente}, M., {Carlstrom}, J.~E., {Joy}, M.~K., {Nagai},
  D., {Reese}, E.~D., \& {Dawson}, K.~S. 2006, \apj, 652, 917

\bibitem[{{Lau} {et~al.}(2009){Lau}, {Kravtsov}, \& {Nagai}}]{lau2009}
{Lau}, E.~T., {Kravtsov}, A.~V., \& {Nagai}, D. 2009, \apj, 705, 1129

\bibitem[{{Mantz} {et~al.}(2008){Mantz}, {Allen}, {Ebeling}, \&
  {Rapetti}}]{mantz2008}
{Mantz}, A., {Allen}, S.~W., {Ebeling}, H., \& {Rapetti}, D. 2008, \mnras, 387,
  1179

\bibitem[{{Mantz} {et~al.}(2010){Mantz}, {Allen}, {Rapetti}, \&
  {Ebeling}}]{mantz2010}
{Mantz}, A., {Allen}, S.~W., {Rapetti}, D., \& {Ebeling}, H. 2010, \mnras, 406,
  1759

\bibitem[{{Marriage} {et~al.}(2011){Marriage}, {Baptiste Juin}, {Lin},
  {Marsden}, {Nolta}, {Partridge}, {Ade}, {Aguirre}, {Amiri}, {Appel},
  {Barrientos}, {Battistelli}, {Bond}, {Brown}, {Burger}, {Chervenak}, {Das},
  {Devlin}, {Dicker}, {Bertrand Doriese}, {Dunkley}, {D{\"u}nner},
  {Essinger-Hileman}, {Fisher}, {Fowler}, {Hajian}, {Halpern}, {Hasselfield},
  {Hern{\'a}ndez-Monteagudo}, {Hilton}, {Hilton}, {Hincks}, {Hlozek},
  {Huffenberger}, {Handel Hughes}, {Hughes}, {Infante}, {Irwin}, {Kaul},
  {Klein}, {Kosowsky}, {Lau}, {Limon}, {Lupton}, {Martocci}, {Mauskopf},
  {Menanteau}, {Moodley}, {Moseley}, {Netterfield}, {Niemack}, {Page},
  {Parker}, {Quintana}, {Reid}, {Sehgal}, {Sherwin}, {Sievers}, {Spergel},
  {Staggs}, {Swetz}, {Switzer}, {Thornton}, {Trac}, {Tucker}, {Warne},
  {Wilson}, {Wollack}, \& {Zhao}}]{marriage2010}
{Marriage}, T.~A. {et~al.} 2011, \apj, 731, 100

\bibitem[{{Maughan} {et~al.}(2008){Maughan}, {Jones}, {Forman}, \& {Van
  Speybroeck}}]{maughan2008}
{Maughan}, B.~J., {Jones}, C., {Forman}, W., \& {Van Speybroeck}, L. 2008,
  \apjs, 174, 117

\bibitem[{{Million} \& {Allen}(2009)}]{million2009}
{Million}, E.~T., \& {Allen}, S.~W. 2009, \mnras, 399, 1307

\bibitem[{{Mroczkowski} {et~al.}(2009){Mroczkowski}, {Bonamente}, {Carlstrom},
  {Culverhouse}, {Greer}, {Hawkins}, {Hennessy}, {Joy}, {Lamb}, {Leitch},
  {Loh}, {Maughan}, {Marrone}, {Miller}, {Muchovej}, {Nagai}, {Pryke}, {Sharp},
  \& {Woody}}]{mroczkowski2009}
{Mroczkowski}, T. {et~al.} 2009, \apj, 694, 1034

\bibitem[{{Muchovej} {et~al.}(2010){Muchovej}, {Leitch}, {Carlstrom},
  {Culverhouse}, {Greer}, {Hawkins}, {Hennessy}, {Joy}, {Lamb}, {Loh},
  {Marrone}, {Miller}, {Mroczkowski}, {Pryke}, {Sharp}, \&
  {Woody}}]{muchovej2010}
{Muchovej}, S. {et~al.} 2010, \apj, 716, 521

\bibitem[{{Muchovej} {et~al.}(2011){Muchovej}, {Leitch}, {Carlstrom},
  {Culverhouse}, {Greer}, {Hawkins}, {Hennessy}, {Joy}, {Lamb}, {Loh},
  {Marrone}, {Miller}, {Mroczkowski}, {Pryke}, {Sharp}, \&
  {Woody}}]{muchovej2011}
---. 2011, \apj, 732, 28

\bibitem[{{Muchovej} {et~al.}(2007){Muchovej}, {Mroczkowski}, {Carlstrom},
  {Cartwright}, {Greer}, {Hennessy}, {Loh}, {Pryke}, {Reddall}, {Runyan},
  {Sharp}, {Hawkins}, {Lamb}, {Woody}, {Joy}, {Leitch}, \&
  {Miller}}]{muchovej2007}
---. 2007, \apj, 663, 708

\bibitem[{{Nagai} {et~al.}(2007){Nagai}, {Vikhlinin}, \&
  {Kravtsov}}]{nagai2007b}
{Nagai}, D., {Vikhlinin}, A., \& {Kravtsov}, A.~V. 2007, \apj, 655, 98

\bibitem[{{Navarro} {et~al.}(1997){Navarro}, {Frenk}, \& {White}}]{navarro1997}
{Navarro}, J.~F., {Frenk}, C.~S., \& {White}, S.~D.~M. 1997, \apj, 490, 493

\bibitem[{{Nevalainen} {et~al.}(2005){Nevalainen}, {Markevitch}, \&
  {Lumb}}]{nevalainen2005}
{Nevalainen}, J., {Markevitch}, M., \& {Lumb}, D. 2005, \apj, 629, 172

\bibitem[{{Nozawa} {et~al.}(2006){Nozawa}, {Itoh}, {Suda}, \&
  {Ohhata}}]{nozawa2006}
{Nozawa}, S., {Itoh}, N., {Suda}, Y., \& {Ohhata}, Y. 2006, Nuovo Cimento B
  Serie, 121, 487

\bibitem[{{Pen}(1997)}]{pen1997}
{Pen}, U. 1997, New Astronomy, 2, 309

\bibitem[{{Peng} \& {Nagai}(2009)}]{peng2009}
{Peng}, F., \& {Nagai}, D. 2009, \apj, 693, 839

\bibitem[{{Percival} {et~al.}(2010){Percival}, {Reid}, {Eisenstein}, {Bahcall},
  {Budavari}, {Frieman}, {Fukugita}, {Gunn}, {Ivezi{\'c}}, {Knapp}, {Kron},
  {Loveday}, {Lupton}, {McKay}, {Meiksin}, {Nichol}, {Pope}, {Schlegel},
  {Schneider}, {Spergel}, {Stoughton}, {Strauss}, {Szalay}, {Tegmark},
  {Vogeley}, {Weinberg}, {York}, \& {Zehavi}}]{percival2010}
{Percival}, W.~J. {et~al.} 2010, \mnras, 401, 2148

\bibitem[{{Press} {et~al.}(1992){Press}, {Teukolsky}, {Vetterling}, \&
  {Flannery}}]{press1992}
{Press}, W.~H., {Teukolsky}, S.~A., {Vetterling}, W.~T., \& {Flannery}, B.~P.
  1992, Numerical Recipes in C. The Art of Scientific Computing (Cambridge:
  Cambridge University Press, 2nd ed.)

\bibitem[{{Reese} {et~al.}(2002){Reese}, {Carlstrom}, {Joy}, {Mohr}, {Grego},
  \& {Holzapfel}}]{reese2002}
{Reese}, E.~D., {Carlstrom}, J.~E., {Joy}, M., {Mohr}, J.~J., {Grego}, L., \&
  {Holzapfel}, W.~L. 2002, \apj, 581, 53

\bibitem[{{Rephaeli}(1978)}]{rephaeli1978}
{Rephaeli}, Y. 1978, \apj, 225, 335

\bibitem[{{Rudy}(1987)}]{rudy1987}
{Rudy}, D.~J. 1987, PhD thesis, California Inst. of Tech., Pasadena.

\bibitem[{{Sarazin} \& {Lieu}(1998)}]{sarazin1998}
{Sarazin}, C.~L., \& {Lieu}, R. 1998, \apjl, 494, L177

\bibitem[{{Sasaki}(1996)}]{sasaki1996}
{Sasaki}, S. 1996, \pasj, 48, L119

\bibitem[{{Saunders} {et~al.}(2003){Saunders}, {Kneissl}, {Grainge},
  {Grainger}, {Jones}, {Maggi}, {Das}, {Edge}, {Lasenby}, {Pooley}, {Miyoshi},
  {Tsuruta}, {Yamashita}, {Tawara}, {Furuzawa}, {Harada}, \&
  {Hatsukade}}]{saunders2003}
{Saunders}, R. {et~al.} 2003, \mnras, 341, 937

\bibitem[{{Sehgal} {et~al.}(2010){Sehgal}, {Trac}, {Acquaviva}, {Ade},
  {Aguirre}, {Amiri}, {Appel}, {Barrientos}, {Battistelli}, {Bond}, {Brown},
  {Burger}, {Chervenak}, {Das}, {Devlin}, {Dicker}, {Bertrand Doriese},
  {Dunkley}, {D{\"u}nner}, {Essinger-Hileman}, {Fisher}, {Fowler}, {Hajian},
  {Halpern}, {Hasselfield}, {Hern{\'a}ndez-Monteagudo}, {Hilton}, {Hilton},
  {Hincks}, {Hlozek}, {Holtz}, {Huffenberger}, {Hughes}, {Hughes}, {Infante},
  {Irwin}, {Jones}, {Baptiste Juin}, {Klein}, {Kosowsky}, {Lau}, {Limon},
  {Lin}, {Lupton}, {Marriage}, {Marsden}, {Martocci}, {Mauskopf}, {Menanteau},
  {Moodley}, {Moseley}, {Netterfield}, {Niemack}, {Nolta}, {Page}, {Parker},
  {Partridge}, {Reid}, {Sherwin}, {Sievers}, {Spergel}, {Staggs}, {Swetz},
  {Switzer}, {Thornton}, {Tucker}, {Warne}, {Wollack}, \& {Zhao}}]{sehgal2010}
{Sehgal}, N. {et~al.} 2010, ArXiv e-prints

\bibitem[{{Snowden} {et~al.}(1998){Snowden}, {Egger}, {Finkbeiner}, {Freyberg},
  \& {Plucinsky}}]{snowden1998}
{Snowden}, S.~L., {Egger}, R., {Finkbeiner}, D.~P., {Freyberg}, M.~J., \&
  {Plucinsky}, P.~P. 1998, \apj, 493, 715

\bibitem[{{Snowden} {et~al.}(1997){Snowden}, {Egger}, {Freyberg}, {McCammon},
  {Plucinsky}, {Sanders}, {Schmitt}, {Truemper}, \& {Voges}}]{snowden1997}
{Snowden}, S.~L. {et~al.} 1997, \apj, 485, 125

\bibitem[{{Sulkanen}(1999)}]{sulkanen1999}
{Sulkanen}, M.~E. 1999, \apj, 522, 59

\bibitem[{{Sunyaev} \& {Zel'dovich}(1972)}]{sunyaev1972}
{Sunyaev}, R.~A., \& {Zel'dovich}, Y.~B. 1972, Comments Astrophys. Space Phys.,
  4, 173

\bibitem[{{Takei} {et~al.}(2008){Takei}, {Miller}, {Bregman}, {Kimura},
  {Ohashi}, {Mitsuda}, {Tamura}, {Yamasaki}, \& {Fujimoto}}]{takei2008}
{Takei}, Y. {et~al.} 2008, \apj, 680, 1049

\bibitem[{{Vanderlinde} {et~al.}(2010){Vanderlinde}, {Crawford}, {de Haan},
  {Dudley}, {Shaw}, {Ade}, {Aird}, {Benson}, {Bleem}, {Brodwin}, {Carlstrom},
  {Chang}, {Crites}, {Desai}, {Dobbs}, {Foley}, {George}, {Gladders}, {Hall},
  {Halverson}, {High}, {Holder}, {Holzapfel}, {Hrubes}, {Joy}, {Keisler},
  {Knox}, {Lee}, {Leitch}, {Loehr}, {Lueker}, {Marrone}, {McMahon}, {Mehl},
  {Meyer}, {Mohr}, {Montroy}, {Ngeow}, {Padin}, {Plagge}, {Pryke}, {Reichardt},
  {Rest}, {Ruel}, {Ruhl}, {Schaffer}, {Shirokoff}, {Song}, {Spieler},
  {Stalder}, {Staniszewski}, {Stark}, {Stubbs}, {van Engelen}, {Vieira},
  {Williamson}, {Yang}, {Zahn}, \& {Zenteno}}]{vanderlinde2010}
{Vanderlinde}, K. {et~al.} 2010, \apj, 722, 1180

\bibitem[{{Vikhlinin} {et~al.}(2006){Vikhlinin}, {Kravtsov}, {Forman}, {Jones},
  {Markevitch}, {Murray}, \& {Van Speybroeck}}]{vikhlinin2006}
{Vikhlinin}, A., {Kravtsov}, A., {Forman}, W., {Jones}, C., {Markevitch}, M.,
  {Murray}, S.~S., \& {Van Speybroeck}, L. 2006, \apj, 640, 691

\bibitem[{{Vikhlinin} {et~al.}(2009){Vikhlinin}, {Kravtsov}, {Burenin},
  {Ebeling}, {Forman}, {Hornstrup}, {Jones}, {Murray}, {Nagai}, {Quintana}, \&
  {Voevodkin}}]{vikhlinin2009}
{Vikhlinin}, A. {et~al.} 2009, \apj, 692, 1060

\bibitem[{{Watkins}(1997)}]{watkins1997}
{Watkins}, R. 1997, \mnras, 292, L59

\bibitem[{{Williamson} {et~al.}(2011){Williamson}, {Vanderlinde}, {Crawford},
  {de Haan}, {Dudley}, {Shaw}, {Ade}, {Aird}, {Benson}, {Bleem}, {Brodwin},
  {Carlstrom}, {Chang}, {Crites}, {Desai}, {Dobbs}, {Foley}, {George},
  {Gladders}, {Hall}, {Halverson}, {High}, {Holder}, {Holzapfel}, {Hrubes},
  {Joy}, {Keisler}, {Knox}, {Lee}, {Leitch}, {Loehr}, {Lueker}, {Marrone},
  {McMahon}, {Mehl}, {Meyer}, {Mohr}, {Montroy}, {Ngeow}, {Padin}, {Plagge},
  {Pryke}, {Reichardt}, {Rest}, {Ruel}, {Ruhl}, {Schaffer}, {Shirokoff},
  {Song}, {Spieler}, {Stalder}, {Staniszewski}, {Stark}, {Stubbs}, {van
  Engelen}, {Vieira}, {Yang}, {Zahn}, \& {Zenteno}}]{williamson2011}
{Williamson}, R. {et~al.} 2011, \apj, submitted

\end{thebibliography}

\begin{table}[h!]
\scriptsize
\centering
\caption{Cluster Observations}
\begin{tabular}{lcccccc|cccc}
\hline
\hline
Cluster & $z$ & $n_{H}$ & $RA$ & $Dec$ &  \multicolumn{2}{c}{\chandra\ Observations} & \multicolumn{4}{c}{SZA Observations}\\
\cline{6-7}
\cline{8-11}
& & $(10^{20} \rm cm^{-2})$ & (J2000) & (J2000) & ObsID & Time  & On-src Time & FWHM & P.A. & rms Noise$^{a}$\\
& & & & & & (ks) & (hrs) & (arcsec) & (deg) & (mJy) \\ 
\hline
A2631  & 0.27 & 3.55 & 23:37:40.1 & +00:16:33 & 3248/11728 &  25.0 & 16.1 & $152 \times 117$ & 17.2 & 0.4  \\
A2204  & 0.15 & 5.67 & 16:32:47.2 & +05:34:32 & 7940 & 72.9  & 19.6 & $157 \times 115$ & -7.7 & 0.4  \\
\hline 
\hline 
\tablenotetext{a}{FWHM (full-width at half maximum of the synthesized beam), P.A.(position
angle of the synthesized beam), and rms noise are for short baselines ($\le 2k\lambda$).}
\label{tab:info} 
\end{tabular} 
\end{table}

\begin{table}[h!]
\footnotesize
\centering
\caption{Radio Sources in Cluster Fields}
\begin{tabular}{lccccccccc}
\hline
\hline
Cluster & \multicolumn{2}{c}{Pointing Center} & \multicolumn{4}{c}{30 GHz Source}\\
\cline{2-3}
\cline{4-7}
& $RA$ & $Dec$ & src & $\Delta \alpha^{a}$ & $\Delta \delta^{a}$ & Flux & FWHM & P.A. & rms Noise$^{b}$\\
& (J2000) & (J2000) & & (arcsec) & (arcsec) & (mJy) & (arcsec) & (deg) & (mJy)\\
\hline
Abell 2631 & 23:37:38.8 & +00:16:06.5  & 1 & 21.3 & 36.5 & 3.7 & $26.5 \times 16.6$ & 42.7 & 0.25\\
           &            &                & 2 & 205.0 & -130.0 & 0.5 \\
Abell 2204 & 16:32:46.88 & +05:34:32.4 & 1 & 0.4 & 1.2 & 7.0 & $21.1 \times 18.4$ & -82.1 & 0.22\\
           &             &               & 2 & -417.8 & -360.1 & 21.6 \\
           &             &               & 3 & 195.0 & -130.1 & 0.7 \\
\hline
\hline
\tablenotetext{a}{$\Delta \alpha$ and $\Delta \delta$ are the offsets from the pointing center.}
\tablenotetext{b}{FWHM, P.A. and rms noise are for long baselines ($> 2k\lambda$).}
\label{tab:radiops}
\end{tabular}
\end{table}

\begin{table}[ht!]
\footnotesize
\centering
\caption{Pressure Normalization Values and Integrated $Y(r_{500})$ Values$^{a}$}
\begin{tabular}{lccc}
\hline
\hline
Cluster & $P_{eo}(SZ)$ & $P_{eo}(X)$  & $P_{eo}(SZ)/P_{eo}(X)$  \\
 & \multicolumn{2}{c}{($10^{-10}~\mathrm{ergs~cm^{-3}}$)} & \\
\hline
Abell 2631        & $1.00\pm^{0.11}_{0.11}$  & $1.21\pm^{0.15}_{0.14}$  & $0.82\pm^{0.09}_{0.09}$ \\
Abell 2204        & $9.90\pm^{0.60}_{0.60}$  & $9.71\pm^{0.47}_{0.47}$  & $1.02\pm^{0.05}_{0.05}$ \\
\hline
& $Y_{\rm sph,SZ} (r_{500})$ & $Y_{\rm sph,X} (r_{500})$  & $Y_{\rm sph,SZ} (r_{500})/Y_{\rm sph,X} (r_{500})$\\
&  \multicolumn{2}{c}{($10^{-11}$)} & \\
\hline
Abell 2631        & $9.13\pm^{1.17}_{1.00}$ & $11.13\pm^{1.52}_{1.34}$ &  $0.82\pm^{0.15}_{0.18}$ \\
Abell 2204        &$44.97\pm^{2.99}_{2.74}$ & $43.93\pm^{3.08}_{2.59}$ & $1.02\pm^{0.10}_{0.09}$\\
\hline
\hline
\tablenotetext{a}{Statistical and \chandra\ calibration systematics are included in the measurement of masses. }
\label{tab:pres}
\end{tabular}
\end{table}

\begin{table}[h!] 
\scriptsize 
\caption{Results from Joint X-ray/SZ Analysis} 
\begin{tabular}{lccccccccc} 
\hline 
\hline 
& \multicolumn{9}{c}{Model Parameters} \\ 
Cluster  & $n_{e0}$ & $r_{s}$ &$n$ & $\beta$ &  $T_{\rm \tiny 0}^{a}$ & $r_{\rm cool}$ & $\alpha$ & $\gamma^{b}$ & \da  \\ 
  & ($10^{-2}$cm$^{-3}$) & (arcsec) & & & (keV) & (arcsec) & & & (Mpc)  \\ 
\hline 
Abell 2631  & $0.78\pm^{0.17}_{0.12}$ & $261.1\pm^{74.6}_{50.2}$ & $9.6\pm^{2.1}_{1.4}$ & $2.0$ & $7.6\pm^{1.7}_{1.6}$ & - & - & -   & $799\pm^{308}_{267}$ \\ 
Abell 2204  & $3.90\pm^{0.25}_{0.19}$ & $22.7\pm^{1.8}_{1.9}$ & $6.9\pm^{1.8}_{1.3}$ & $1.37\pm^{0.10}_{0.08}$ & $14.9\pm^{1.6}_{0.9}$ & $20.0\pm^{0.7}_{0.7}$  & $0.17\pm0.01$  & $2.0$  & $575\pm^{47}_{56}$ \\ 
\hline 
\hline 
& \multicolumn{8}{c}{Cluster Masses$^{c}$} \\ 
& \multicolumn{4}{c}{Masses Evaluated At $\Delta$= 2500} & \multicolumn{4}{c}{Masses Evaluated At $\Delta$=500} \\ 
 & \multicolumn{4}{c}{\hrulefill} & \multicolumn{4}{c}{\hrulefill}\\
Cluster & $r_{\Delta}$ & $\Mgas\ $ & $\Mtot\ $ & $\fgas$ & $r_{\Delta}$ & $\Mgas\ $ & $\Mtot\ $ & $\fgas$\\ 
& ('') & $(10^{13} M_{\odot})$ & $(10^{14} M_{\odot})$ & & ('') & $(10^{13} M_{\odot})$ & $(10^{14} M_{\odot})$ & \\ 
\hline 
Abell 2631  & $103.2\pm^{18.4}_{18.3}$ & $1.71\pm^{2.40}_{1.16}$ & $1.37\pm^{1.16}_{0.65}$ & $0.124\pm^{0.081}_{0.060}$ 
                  & $289.3\pm^{44.0}_{41.6}$ & $8.35\pm^{10.77}_{5.33}$ & $6.16\pm^{4.35}_{2.62}$ & $0.131\pm^{0.103}_{0.065}$\\ 
Abell 2204  & $231.3\pm^{9.0}_{8.3}$ & $5.03\pm^{1.08}_{1.07}$ & $4.09\pm^{0.39}_{0.34}$ & $0.122\pm^{0.019}_{0.018}$ 
                  & $492.6\pm^{23.3}_{20.7}$ & $13.09\pm^{2.96}_{2.94}$ & $7.95\pm^{1.20}_{0.92}$ & $0.161\pm^{0.026}_{0.021}$ \\ 
\hline 
\hline 
\tablenotetext{a}{The reader is cautioned that $T_{\rm \tiny 0}$ is not a global temperature, 
but rather a model parameter in Equation \ref{eq:temp}.} 
\tablenotetext{b}{The parameter $\gamma$ is fixed in the model.}
\tablenotetext{c}{Statistical and \chandra\ calibration systematics are included in the measurements. }
\label{tab:joint} 
\end{tabular} 
\end{table}

\begin{table}[h!]
\scriptsize
\centering
\caption{Results from X-ray Analysis}
\begin{tabular}{lccccccccc}
\hline
\hline
& \multicolumn{9}{c}{Model Parameters} \\
Cluster  & $n_{e0}$ & $r_{s}$ &$n$ & $\beta$ &  $T_{\rm \tiny 0}^{a}$ & $r_{\rm cool}$ & $\alpha$ & $\gamma^{b}$ & \da$^{b}$ \\
  & ($10^{-2}$cm$^{-3}$) & (arcsec) & & & (keV) & (arcsec) & & & (Mpc)  \\
\hline
Abell 2631  & $0.7r\pm^{0.03}_{0.03}$ & $248.0\pm^{67.6}_{47.0}$ & $9.27\pm^{1.91}_{1.34}$ & $2.0$ & $8.3\pm^{0.7}_{0.7}$ & - & - & -   & $840.0$ \\
Abell 2204  & $4.12\pm^{0.23}_{0.25}$ & $22.5\pm^{1.7}_{1.5}$ & $6.76\pm^{1.19}_{0.82}$ & $1.38\pm^{0.06}_{0.06}$ & $15.0\pm^{0.8}_{0.8}$ & $20.0\pm^{0.7}_{0.6}$  & $0.17\pm^{0.08}_{0.09}$  & $2.0$  & $526.0$ \\
\hline
\hline
& \multicolumn{8}{c}{Cluster Masses$^{c}$} \\
& \multicolumn{4}{c}{Masses Evaluated At $\Delta$= 2500} & \multicolumn{4}{c}{Masses Evaluated At $\Delta$=500} \\
     & \multicolumn{4}{c}{\hrulefill} & \multicolumn{4}{c}{\hrulefill}\\
Cluster & $r_{\Delta}$ & $\Mgas\ $ & $\Mtot\ $ & $\fgas$ & $r_{\Delta}$ & $\Mgas\ $ & $\Mtot\ $ & $\fgas$\\
& ('') & $(10^{13} M_{\odot})$ & $(10^{14} M_{\odot})$ & & ('') & $(10^{13} M_{\odot})$ & $(10^{14} M_{\odot})$ & \\
\hline
Abell 2631 & $113.9\pm^{6.6}_{7.0}$ & $2.45\pm^{0.26}_{0.27}$ & $1.98\pm^{0.37}_{0.35}$ & $0.124\pm^{0.010}_{0.008}$ 
                  & $309.9\pm^{20.7}_{18.3}$ & $10.34\pm^{0.50}_{0.49}$ & $8.00\pm^{1.72}_{1.33}$ & $0.129\pm^{0.019}_{0.018}$\\
Abell 2204  & $234.4\pm^{3.8}_{3.9}$ & $4.12\pm^{0.09}_{0.09}$ & $3.75\pm^{0.19}_{0.18}$ & $0.110\pm^{0.003}_{0.003}$ 
                  & $499.0\pm^{11.8}_{12.5}$ & $10.64\pm^{0.27}_{0.29}$ & $7.27\pm^{0.53}_{0.53}$ & $0.146\pm^{0.007}_{0.007}$ \\
\hline
\hline
\tablenotetext{a}{The reader is cautioned that $T_{\rm \tiny 0}$ is not a global temperature,
but rather a model parameter in Equation \ref{eq:temp}.}
\tablenotetext{b}{Parameters $\gamma$ and $D_A$ are fixed in the model.}
\tablenotetext{c}{Statistical and \chandra\ calibration systematics are included in the measurement of masses. }
\label{tab:xray}
\end{tabular}
\end{table}

\begin{table}[h!]
\footnotesize
\centering
\caption{Sources of Uncertainty }
\begin{tabular}{lccc|ccc}
\hline
\hline
& \multicolumn{3}{c}{$r_{2500}$} \vline & \multicolumn{3}{c}{$r_{500}$} \\
\cline{2-4}
\cline{5-7}
& Effect on \fgas & \multicolumn{2}{c}{ Effect on Pressure} \vline & Effect on \fgas & \multicolumn{2}{c}{Effect on Pressure}\\
\cline{3-4}
\cline{6-7}
Source & (\%) & SZ (\%) & X-ray (\%) & (\%) & SZ (\%) & X-ray (\%) \\
\hline
Kinetic SZ effect & $\pm4$  & $\pm4$ & ...&  $\pm4$  & $\pm4$ & ...\\
Radio Point Sources & $\pm{2}$ & $\pm{1}$ & ...&  $\pm{2}$ & $\pm{1}$ & ...\\
Asphericity & $\pm20$ & $\pm10$ & $\pm10$ & $\pm20$ & $\pm10$ & $\pm10$ \\
X-ray background & $\pm{5}$ & ... & $\pm{1}$ & $\pm{9}$ & ... & $\pm{2}$\\
SZA calibration & $\pm{10}$ & $\pm{6}$ & ... & $\pm{10}$ & $\pm{6}$ & ...\\
Hydrostatic Equilibrium$^{a}$ & -9 & ... & ...& -11 & ... & ...\\
Model Assumptions & $\pm10$ & $\pm3$ & $\pm3$& $\pm10$ & $\pm5$ & $\pm5$\\
\underline{Helium Sedimentation} & \underline{+10} & \underline{...} & \underline{-4}& \underline{+5} & \underline{...} & \underline{-2}\\
Total Systematic & $\pm{28}$ & $\pm{13}$ & $\pm{11}$ & $\pm^{27}_{29}$ & $\pm{14}$ & $\pm{12}$   \\
\hline
\hline
\underline{\chandra\ calibration uncertainties$^{b}$}\\
Surface Brightness & $\pm10$\\
Temperature & $\pm1$\\
\hline
\hline
\tablenotetext{a}{Uncertainty is theoretically motivated by \cite{lau2009} as discussed in Section 4.3.4.}
\tablenotetext{b}{These systematic uncertainites are added to the data prior to the fit, and their effect is included in the derived masses and pressure at all radii.}
\label{tab:unc}
\end{tabular}
\end{table}

\clearpage
\section*{Appendix: MCMC Reparameterization 
Using a Singular Value Decomposition Method}

Correlation among model parameters is a common feature of analytic
models such as the beta model \citep{cavaliere1976}, the \cite{vikhlinin2006} model,
and the \cite{bulbul2010} model used in this paper.
The Monte Carlo Markov chain (MCMC) method for the analysis of X-ray and
SZE data described in \cite{bonamente2004} accounts for this correlation, and
therefore correlation is not an issue when evaluating integrated quantities
such as masses and $Y$ values, and their uncertainties.
Strong parameter correlation, however, may cause the MCMC to be inefficient
in its sampling of parameter space \citep[see, e.g.,][page 90]{gilks1996},
requiring long chains with low acceptance rate because of the poor mixing.
A common solution is 
the use of a singular value decomposition \citep[SVD, e.g.,][]{press1992} 
to perform a linear transformation of the
parameters
to reduce the
correlation among model parameters, and increase of the rate of acceptance 
in  the Monte Carlo Markov chain.
For the X-ray analysis of the \chandra\
data of Abell~2631 shown in Table~\ref{tab:xray}, the four model parameters ($n_{e0}$, $r_s$, $n$
and $\beta$) are transformed into four SVD parameters ($svd_0$ through $svd_3$), and the usual Metropolis-Hastings
MCMC is applied to the SVD parameters.
The accepted parameters are then
transformed back to the original \cite{bulbul2010} model parameters, for which
we calculate integrated quantities.

The effect of the reparameterization
is shown in Figures~\ref{fig:a2631-correlation-1} and \ref{fig:a2631-correlation-2}.
The strong correlation present between certain pairs of parameters, especially
$r_s$ and $n$, is absent from the SVD parameters.
\begin{figure}[!h]
\centering
\includegraphics[angle=-90,width=5in]{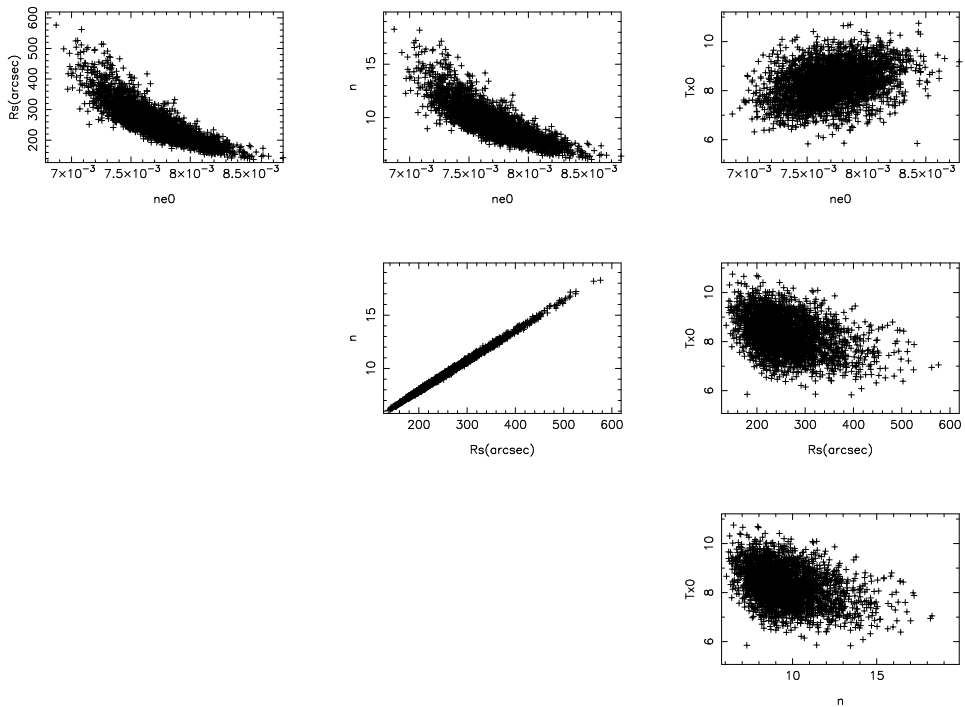}
\caption{Scatter plot for parameters of the \cite{bulbul2010} model applied to the
X-ray analysis of Abell~2631. Notice the strong correlation among some of the parameters,
especially $n$ and $r_s$. For clarity, only every 100-th parameter in the chain is plotted.}
\label{fig:a2631-correlation-1}
\end{figure}
\begin{figure}[!h]
\centering
\includegraphics[angle=-90,width=5in]{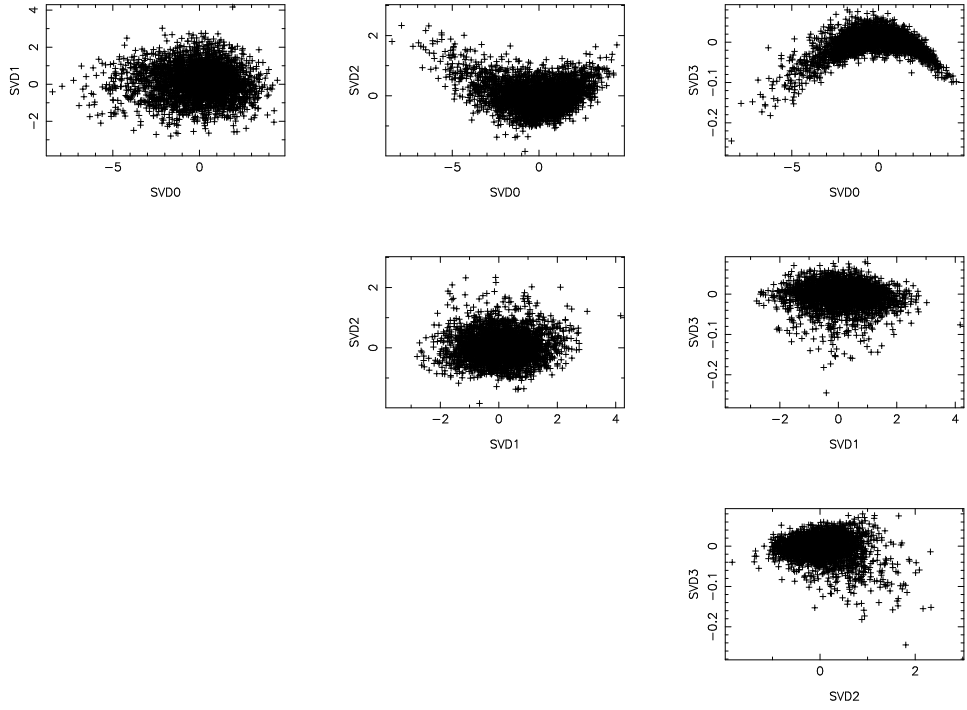}
\caption{Scatter plot for the SVD parameters of the same chain as in 
Figure~\ref{fig:a2631-correlation-1}. }
\label{fig:a2631-correlation-2}
\end{figure}
The values of the correlation coefficients for the original parameters and the SVD parameters
are shown in Table~\ref{tab:covariance}. With this reparameterization, we obtain
an acceptance rate of approximately 30\%, which is a factor of few higher than the
typical acceptance rate obtained using the original parameters. 

\begin{table}[ht!]
\centering
\caption{Correlation Coefficients for the X-ray Analysis of Abell~2631}
\begin{tabular}{lcccc}
\hline
\hline
Parameter & $n_{eo}$ & $r_{s}$ & $n$ & $T_{xo}$  \\
\hline
$n_{eo}$ & \nodata &  -0.85 & -0.82 & 0.31  \\
$r_{s}$  &      &  \nodata  & 1.00  & -0.37 \\
$n$      &      &        &  \nodata & -0.35 \\
$T_{xo}$ &      &        &       & \nodata  \\
\hline
\hline
Parameter & $svd_0$ & $svd_1$ & $svd_1$ & $svd_3$  \\
\hline
$svd_0$ & \nodata &  -0.01 & -0.14 & 0.21  \\
$svd_1$ &      &  \nodata  & 0.02  & -0.04 \\
$svd_2$ &      &        &  \nodata & -0.15 \\
$svd_3$ &      &        &       & \nodata  \\
\hline
\hline
\label{tab:covariance}
\end{tabular}
\end{table}

\end{document}